\documentclass[preprint,nofootinbib]{revtex4-1}
\usepackage{times,graphicx,amsfonts,amsmath,amssymb,color}
\usepackage{slashed}

\begin{document}

\title{An Introduction to Particle Dark Matter\footnote{This article is based on a mini-course of 5 lectures delivered by Professor S. Profumo at the \textit{Third Jos\'e Pl\'i­nio Baptista School on Cosmology}, in September 2016. See \url{http://www.cosmo-ufes.org/jpbcosmo3.html}.}}

\author{S. Profumo$^1$}
\email{profumo@ucsc.edu}
\affiliation{$^1$ Santa Cruz Institute for Particle Physics, Department of Physics, University of California, Santa Cruz, 1156 High Street, Santa Cruz, CA 95064, USA.}
\author{L. Giani$^{2,3,4}$ and O. F. Piattella$^{2,3,4}$}
\email{oliver.piattella@cosmo-ufes.org}
\affiliation{$^{2}$ Department of Physics, Universidade Federal do Esp\'irito Santo, Avenida Fernando Ferrari 514, 29075-910 Vit\'oria, Esp\'irito Santo, Brazil.}
\affiliation{$^{3}$ N\'ucleo Cosmo-ufes and PPGCosmo, Universidade Federal do Esp\'irito Santo, Avenida Fernando Ferrari 514, 29075-910 Vit\'oria, Esp\'irito Santo, Brazil.}
\affiliation{$^{4}$ Institut f\"ur Theoretische Physik, Ruprecht-Karls-Universit\"at Heidelberg, Philosophenweg 16, 69120 Heidelberg, Germany.}

\begin{abstract}
	We review the features of Dark Matter as a particle, presenting some old and new instructive models, and looking for their physical implications in the early universe and in the process of structure formation. We also present a schematic of Dark Matter searches and introduce the most promising candidates to the role of Dark Matter particle.
\end{abstract}

\maketitle


\section{Introduction}

In this review we employ units $\hbar = c = k_B = 1$ units, except where otherwise stated.


\subsection{Why do we need Dark Matter?}

Cosmic Microwave Background (CMB) observations strongly suggest that the universe is, at large scales, nearly spatially flat \cite{Aghanim:2018eyx}, giving an observed value for the present energy density $\rho_0$ close to its critical value:
\begin{equation}\label{rho0}
    \rho_0 \approx \rho_{\rm crit} \equiv \frac{3H_{0}^2}{8\pi G_{N}} \approx 10^{-29} \mbox{ g/cm}^3\;,
\end{equation}
corresponding to an average density of 10 protons per cubic meter; in Eq.~\eqref{rho0} $H_0 \approx 70$ km/s/Mpc is the value of the Hubble factor today (i.e. the Hubble constant) and $G_N \approx 6.67\times 10^{-11}$ N m$^2$/kg$^2$ is Newton's gravitational constant.

This energy density is made up of various contributions \cite{Aghanim:2018eyx}. Mostly (about 70\%) of Dark Energy (DE), which is responsible for the accelerated expansion of the universe, followed by Dark Matter (DM, about 26\%), which is responsible for the gravitational collapse of ordinary matter (usually called ``baryons'' and which amounts to the remaining 4\%) and eventually for the formation of the structures that we observe in the sky.

There are several probes able to weigh the Dark Energy (DE) contribution to this average density: for example, CMB data \cite{Aghanim:2018eyx}, the distribution of galaxies on large scales, also known as the Large Scale Structure (LSS) of the universe \cite{Abbott:2017wau} and a byproduct of the latter, the so-called Baryon Acoustic Oscillations (BAO) \cite{Alam:2016hwk}. Moreover, we can also discriminate the contribution of ordinary baryonic matter and DM, via e.g. Big Bang Nucleosynthesis (BBN) and again CMB observations, see e.g. Refs. \cite{Dodelson:2003ft, Piattella:2018hvi}. In fact, we are able to determine the amount of DM in the universe, which is the following:
\begin{equation}\label{OmegaDM}
    \Omega_{\rm DM} \equiv \frac{\rho_{\rm DM,0}}{\rho_0} \approx 0.26\;,
\end{equation}
where $\rho_{\rm DM,0}$ is the DM density evaluated today (usually a $0$ subscript means, in cosmology, that the corresponding quantity has to be evaluated at the present time, i.e. at the age of the universe $t_0$). The number in Eq.~\eqref{OmegaDM} can be rewritten in units more suitable to astrophysics and particle physics as follows:
\begin{equation}
    \rho_{\rm DM,0} \approx 10^{10} \frac{M_{\odot}}{\mbox{Mpc}^3} \approx 10^{-6} \frac{\rm GeV}{\rm cm^3}\;,
\end{equation}
where $M_{\odot} \approx  2 \times 10^{30} \mbox{ kg}$ is the mass of the Sun. 

Note that $\rho_{\rm DM,0}$ is $10^5$ times less dense than the average density in clusters of galaxies and $10^6$ less dense than galaxies. Thus, we can conclude, from the point of view of perturbations about a homogeneous and isotropic background, that the universe is in a highly nonlinear regime. 

Dark Matter is a crucial ingredient, necessary in order to produce the large inhomogeneities that form structures in the universe. Indeed, looking at the CMB sky, we measure very small temperature fluctuations, proportional to the baryonic density fluctuations, which we infer being of order:
\begin{equation}
    \delta_{\rm b} \equiv \frac{\delta \rho_{\rm b}}{\rho_{\rm b}} \lesssim 10^{-4}\;, \qquad (\mbox{at recombination, i.e. for } z \approx 1100)\;,
\end{equation}
where $\delta_{\rm b}$ is the baryonic density contrast, i.e. the ratio of density fluctuations in the baryonic density and its background value. This number is important because we also know that sub-horizon pressureless matter (as baryons are) overdensities in the linear regime grow linearly with the scale factor in the matter-dominated epoch of the evolution of the universe. Therefore, from the recombination epoch at $z_{\rm rec} \approx 1100$ baryon fluctuations, which are finally free to grow, have not enough time for entering the nonlinear regime, but they would be of order $10^{-1}$ today, so still linear. This issue is solved if a different ``dark'' species, which decoupled from photons much earlier than baryons did, is present, giving time for its perturbations to grow enough in order to trigger, at the recombination epoch, a faster collapse of baryonic fluctuations and thus eventually making structure formation possible.

For this reason, it is in general difficult to explain consistently structure formation without recurring to DM. See e.g. \cite{Dodelson:2011qv} for the case against MOdified Newtonian Dynamics (MOND) \cite{Milgrom:2009an} and its covariant version TeVeS \cite{Bekenstein:1984tv}.


\subsection{Microscopic features of Dark Matter}

We can get some clues about the microscopic nature of DM by trying to answer the following questions:

\subsubsection{Is Dark Matter actually dark?} It is generally believed that DM is not bright, in the sense that DM particles do not interact electromagnetically and so are not able to scatter photons; in other words they do not possess an electric charge. However, observation does not exclude DM particles to have indeed a tiny electric charge \cite{Munoz:2018pzp}. 

\subsubsection{Is Dark Matter collisionless?} In order for collisional effects in the DM sector to be relevant, the mean free path $\lambda_{\rm MFP}$ of a DM particle must be larger than the typical size of a cluster of galaxies, i.e. $\approx 1$ Mpc, which has a density of order $\rho_{\rm cl} \approx 1$ ${\rm GeV}/{\rm cm^3}$. Therefore, we get, for a DM particle with mass $m = 1$ GeV, the following result for the cross section $\sigma$:
\begin{equation}
        \lambda_{\rm MFP} = \frac{1}{\sigma\rho_{\rm cl}/m} > 1 \mbox{ Mpc} \quad \Rightarrow \quad  \sigma/m \lesssim {\rm cm^2}/{\rm g} \approx {\rm barn}/{\rm GeV}\;,
\end{equation}
which is a cross section whose value is typical of the strong interaction. This suggests that DM might not be completely collisionless and makes models of Self-interacting Dark Matter (SIDM) appealing. It is worth also to mention scenarios of Partially Interacting DM, where only a subdominant component of the total DM contribution is self-interacting \cite{Fan:2013yva, Fan:2013bea}.
    
\subsubsection{Is Dark Matter classical?} We know that DM must be gravitationally bounded on scales at least as large as the size of a dwarf spheroidal (dSph) galaxy. We can use this size in order to derive an upper bound on the de Broglie wavelength $\lambda_{\rm dB}$ of a DM particle and, consequently, a lower bound on its mass. Indeed, consider a DM particle of mass $m$ with velocity $v \approx 100$ km/s. Using the Planck's constant value (which we recover for the moment) $h_{\rm Pl} \approx 4 \times 10^{-15}$ eV$\cdot$s, and the fact that the particle is non-relativistic, the de Broglie wavelength $\lambda_{\rm dB} = h_{\rm Pl}/(mv)$ is given by:
\begin{equation}
	\lambda_{\rm dB} \approx \frac{4\times 10^{-15}}{10^2} \frac{\mbox{eV}}{m}\frac{\mbox{s}^2}{\mbox{km}} \approx 4 \mbox{ mm }\frac{\mbox{ \rm eV}}{m}\;.
\end{equation}
This result means that in order to have $\lambda_{\rm dB} < 1$ kpc $\approx 3\times 10^{21}$ cm we need a mass 
\begin{equation}\label{mbound1}
	m > 10^{-22}\mbox{ eV}\;,
\end{equation}
which is not a very stringent lower bound. 

On the other hand, we know that quantum effects can drastically change such a lower bound e.g. if DM is a fermion, because in this case its phase space density is bounded by the so-called Pauli blocking effect (i.e. two fermions cannot occupy the same quantum state, because of Pauli's exclusion principle). Using the simplest model of a singular isothermal sphere with $\rho_{\rm dSph} = \sigma^2_{\rm dSph}/(2\pi G_Nr^2)$, with $r$ radius of the sphere and $\sigma^2_{\rm dSph}$ a constant velocity dispersion one obtains the so-called \textit{Tremaine-Gunn limit} \cite{Tremaine:1979we}:
\begin{equation}\label{TremaineGunnbound}
    m^4 > \frac{h_{\rm Pl}^3}{(2\pi)^{5/2}gG_N\sigma_{\rm dSph}r^2} \approx \left(38\mbox{ eV}\right)^4\left(\frac{100\mbox{ km/s}}{\sigma_{\rm dSph}}\right)\left(\frac{10\mbox{ kpc}}{r}\right)^2g^{-1}\;,
\end{equation}
where $g$ is the number of spin states of the DM particle. The difference of 23 orders of magnitude in the above two bounds in Eqs.~\eqref{mbound1} and \eqref{TremaineGunnbound} for the DM particle mass teaches us how important quantum effects might be in the description of DM.


\subsubsection{Is Dark Matter a fluid?} It is possible to describe DM as a fluid? In this case, we can forget about the particle nature of DM and simply describe it by using its density and velocity flow. This is also called \textit{fluid approximation}. We can get some constraints on the coarse graining of such a fluid, i.e. on the scale down to which the fluid approximation works, by requiring that it must not disrupt e.g. clusters of stars. In fact, if the grains constituting this fluid are too big, e.g. black holes, they must have effects on bound compact systems such as globular clusters. In Ref. \cite{1985ApJ} it is showed that this would be the case for grains with mass $m \geq 10^6 M_\odot $. 

The estimate energy exchanged in collisions among globular clusters and black holes, using the impulse approximation according to which the interactions are instantaneous (as if the bodies were rigid spheres), demand that the energy must be smaller than the gravitational binding energy of the bound compact system (roughly proportional to $G_NM^2/R$, where $M$ is the mass of the system and $R$ its size), giving a mass of order $m \lesssim 10^3 M_\odot \approx 10^{70} \mbox{ eV}$, see e.g. Ref.~\cite{Goerdt:2006hp}.

In conclusion, from the above phenomenological considerations we can list the following properties of particle DM:
\begin{description}
\item[Mass:] There are more than 90 orders of magnitude of mass range available for bosons; 70 for fermions.
\item[Interactions:] A self-interaction is possible, if of the order at most of the strong interaction, so of the order of the MeV. It is in principle also possible to have interactions with ordinary Standard Model particles, as long as such interactions do not involve emission of photons, otherwise DM halos would shine and be visible. Particularly promising as DM candidates are massive particles that interact only via weak interactions, the so-called WIMPs (Weakly Interacting Massive Particle).
\item[Abundance:] It has to be enough in order to satisfy the observational constraint $\Omega_{\rm DM} \approx 0.26$. As we shall see later, this is achieved naturally for WIMPs, and such occurrence is usually referred to as the \textit{WIMP miracle}.
\end{description}


\subsection{Observational constraints on Dark Matter interactions} 

While only loose model-independent constraints exist on DM self-interaction and on DM interaction with Standard Model particles, a broad portfolio of experimental probes has been deployed over the years with the aim of searching for signals coming from specific DM particle models. The best-developed searches focus on WIMPs, and seek to capture the elusive signals from elastic scattering of WIMPs off nucleons (direct detection), signals of WIMP pair-annihilation or decay (indirect detection), and the pair-production of WIMPs at colliders in conjunction with detectable Standard Model particles.

So far, while several signals have been claimed to be possibly associated with different DM candidates, including WIMPs, no conclusive detection has been agreed upon. Null results from direct detection experiments at present constrain proton-WIMP interactions to be below the level of a cross section of $10^{-47}\ {\rm cm}^2$ for a mass around the weak scale (i.e. about 100 GeV); the absence of emission of gamma-rays from DM-dominated (this DM-domination is inferred of course by gravitational effects) nearby structures such as dSph galaxies implies a pair-annihilation cross section times relative velocity smaller than the required one for thermal production in $s$-wave annihilation (which means that the thermal averaged cross section is velocity-independent) for masses between around 10 and 100 GeV, with weaker limits at larger masses. Finally, no conclusive result from colliders has been reported, with resulting limits which depend upon the specific WIMP model.

We recommend to the reader the recent review on WIMP searches given in Ref.~\cite{Arcadi:2017kky} for further details.


\section{Thermal decoupling}

Thermal decoupling provides a successful framework able to explain the origin of species in the early universe. Built on a synergy among Statistical Mechanics, Particle Physics and General Relativity, it allows to make highly accurate testable predictions.

The key idea of thermal decoupling is based on the interplay between the rate $\Gamma$ of a certain interaction process and the expansion of the universe, described by the Hubble factor $H$, both time-dependent. When $\Gamma \gg H$, the interaction is very efficient and, if we are dealing with annihilation and pair productions, they are in equilibrium (more precisely, chemical equilibrium). However, because of the expansion of the universe, in general $\Gamma$ becomes at some time smaller than the Hubble factor $H$. At this point the expansion of the universe thwarts the interaction, making it inefficient. Taking again the example of the annihilation, when $\Gamma \approx H$ it stops to be efficient and the species which was previously annihilating is said to \textit{freeze-out}. Species that were originally in thermal equilibrium with the primordial plasma and then freeze-out are said to be \textit{thermally produced}. 

The reaction rate $\Gamma$ for a given interaction is specified by the relevant cross section $\sigma$, the number density $n$ of the particle species and the relative velocity $v$:
\begin{equation}\label{Gammadef}
	\Gamma = n \sigma v\;.
\end{equation}
We know from Statistical Mechanics that the number density at thermal equilibrium at temperature $T$ is given by (see Ref.~\cite{Piattella:2018hvi} for a detailed derivation of these formulae):
\begin{eqnarray}
	\label{nrel} n_{\rm rel} \propto T^{3} \quad \mbox{for } m \ll T\;,\\
	\label{nnonrel} n_{\rm non-rel} \propto \left(mT\right)^{3/2}\exp\left(-m/T\right) \quad \mbox{for } m \gg T\;,
\end{eqnarray}
where the subscripts refer to the fact that particles of a given species in thermal equilibrium are mostly relativistic or non-relativistic depending whether the temperature of the thermal bath is much larger or much smaller than the mass of the particle species under consideration.

The Hubble rate $H$ is given by the Friedmann equation:
\begin{equation}
    H^2 = \frac{8\pi G_N}{3}\rho\;.
\end{equation}
Using Bose-Einstein statistics, one can show that:
\begin{equation}
	\rho_{\rm bos} = \frac{\pi^2}{30}gT^4\;,
\end{equation}
where $g$ is again the number of spin states of the particle boson species. For fermions, the above result still holds true if multiplied by a factor 7/8. We can write in a compact way the radiation density as follows:
\begin{equation}
    \rho_{\rm rad} = \frac{\pi^2}{30}g_{\rm eff}T^4\;, 
\end{equation}
where $g_{\rm eff}$ is an effective total number of spin states of the relativistic particle species, taking into account the factor 7/8 for relativistic fermions. Note that $g_{\rm eff}$ varies with time because massive species becomes non-relativistic as long as the thermal bath cools down, but, after DM and the heaviest particles have decoupled from the primordial plasma, $g_{\rm eff}$ remains basically of order 10. Therefore, we can finally combine our knowledge of Cosmology (i.e. the Friedmann equation) and Statistical Mechanics and obtain:
 \begin{equation}
    H \approx \frac{T^2}{M_{P}}\;,
\end{equation}
where $M_P \approx 10^{19}$ GeV is the Planck mass.


\subsection{Thermal relics}

Relics are particles left over when their annihilation reactions are no more efficient, i.e. when their abundance attains a constant (comoving, i.e. not taking into account the standard dilution due to the expansion of the universe) value. If a species freezes-out when relativistic then it is dubbed \textit{hot thermal relic}, whereas it is a \textit{cold thermal relic} if it freezes-out when non-relativistic.


\subsubsection{Hot thermal relics}

A simple calculation of a hot thermal relic abundance can be carried out e.g. for neutrinos, which constitute the cosmic neutrino background (C$\nu$B) once decoupled. To this purpose, let us consider the interactions:
\begin{equation}
    \nu + \bar{\nu} \leftrightarrow f + \bar{f}\;,
\end{equation}
of neutrino annihilation and neutrino pair production through fermion-antifermion annihilation. Weak interaction rule this process, hence the cross section is given by:
\begin{equation}
	\sigma \simeq G_F^2T_\nu^2\; , 
\end{equation}
where $G_F \approx 10^{-5}$ GeV$^{-2}$ is the Fermi coupling constant and $T_\nu$ is the neutrino thermal bath temperature. The freeze-out condition $\Gamma = H$ gives us, using Eq.~\eqref{Gammadef}:
\begin{equation}
	n_\nu(T_{\nu,\rm freeze-out})\sigma(T_{\nu,\rm freeze-out}) = H(T_{\nu,\rm freeze-out})\;,
\end{equation}
where we have taken into account that neutrinos are relativistic (i.e. $v = 1$, since we are working in natural $c = \hbar = 1$ units). For the same reason, from Eq.~\eqref{nrel} we have that $n_\nu \sim T_\nu^3$ and thus combining the above relations we get:
\begin{equation}
    T_{\nu,\rm freeze-out} \simeq (G_F^2M_P)^{-1/3} \approx 1 \mbox{ MeV}\;.
\end{equation}
For the above estimate we have used Fermi effective Lagrangian of the weak interaction and the fact that the particle is relativistic. The first approximation is reasonable because $m_{\nu} \ll m_{W}$, with $m_W$ mass of the W-boson, so that the interaction can be considered pointlike, and $T_{\nu} \gg m_{\nu}$ is a condition which holds essentially through all the thermal history of the universe (maybe up to recombination, if we consider $m_\nu = 0.1$ eV). 
 
So far we have obtained the freeze-out temperature of the relic neutrinos background, but how do we now evaluate their abundance? Let us introduce the quantity $Y_\nu = n_\nu/s$, where $s$ is the entropy density. Since $S = sa^3$ is conserved then $Y_\nu \propto n_\nu a^3$ is a constant. Therefore:
\begin{equation}
	Y_{\nu}(T_{\nu,0}) = Y_{\nu}(T_{\nu,\rm freeze-out})\;,
\end{equation}
where $T_{\nu,0}$ is the neutrino temperature today. In particular, we can write:
\begin{equation}
	Y_{\nu}(T_{\nu,0}) = \frac{n_{\nu,0}}{s_{0}} = Y_{\nu}(T_{\nu,\rm freeze-out}) = \frac{n_\nu\left(T_{\nu,\rm freeze-out}\right)}{s\left(T_{\nu,\rm freeze-out}\right)}\;,
\end{equation}
and if we assume neutrinos to be non-relativistic today, so that $\rho_{\nu,0} = n_{\nu,0}m_\nu$, we obtain:
\begin{equation}
    \rho_{\nu,0} = m_{\nu}Y_{\nu}(T_{\nu,\rm freeze-out})s_{0}\;,
\end{equation}
and from this relation one determines the so-called \textit{Cowsik-McClelland} bound \cite{PhysRevLett.29.669}:
\begin{equation}
    \Omega_{\nu} \equiv \frac{\rho_{\nu,0}}{\rho_0} \approx \frac{m_{\nu}}{94\;h^2\mbox{ eV}} \le 1\;,
\end{equation}
where $h$ is such that $H_0 = 100\;h\mbox{ km/s/Mpc}$.


\subsubsection{Cold thermal relics}

Let us see now how to proceed in the case of a cold relic of mass $m_\chi$, for which we must use the non-relativistic expression \eqref{nnonrel} for the number density:
\begin{equation}
    n_\chi \simeq \left(m_{\chi}T \right)^{3/2}\exp\left(-m_{\chi}/T\right)\;.
\end{equation}
Combining the latter with the freeze-out condition:
\begin{equation}
    n_{\chi, \rm freeze-out} \simeq \frac{T_{\chi,\rm freeze-out}^2}{M_P \sigma}\;,
\end{equation}
and defining the new variable $x \equiv m_{\chi}/T_{\chi}$, we easily obtain:
\begin{equation}
    \sqrt{x}e^{-x} \simeq \frac{1}{M_{P} \sigma m_{\chi}}\;.
\end{equation}
For the case of WIMPs  we have $m_{\chi} \approx 10^2$ GeV, and $\sigma \simeq G_{F}^{2}m_{\chi}^{2} $, and the solution of the above equation becomes:
\begin{equation}
    x_{\rm freeze-out} \simeq \frac{m_{\chi}}{T_{\chi,\rm freeze-out}} \approx 35\;.
\end{equation}
From this, we can calculate the thermal relic density of WIMPs. As first step, let us write:
\begin{equation}\label{Omegachi0}
    \Omega_{\chi} = \frac{m_{\chi}n_{\chi,0}}{\rho_{0}} = \frac{m_{\chi}T_{0}^{3}}{\rho_{0}} \frac{n_0}{T_{0}^3}\;.
\end{equation}
It is important to stress that in the above computation we are assuming a DM-symmetric universe,\footnote{This is not the case in the so-called \textit{Asymmetric WIMPs} models, see for example Ref.~\cite{Graesser:2011wi}.}, i.e. there is roughly the same amount of DM and anti-DM. Assuming again conservation of the entropy density $s$, so that $aT =$ constant, we can use the relation
\begin{equation}\label{relnchiT03}
	\frac{n_{\chi,0}}{T_{\chi,0}^3} \approx \frac{n_{\chi,\rm freeze-out}}{T^{3}_{\chi,\rm freeze-out}}\;,
\end{equation}
we finally obtain:
\begin{equation}\label{Omegachieq}
    \left(\frac{\Omega_{\chi}}{0.2}\right) \approx \frac{x_{\rm freeze-out}}{20}\left(\frac{10^{-8} \mbox{ GeV}^{-2}}{\sigma}\right)\;.
\end{equation}
In the above estimate we have neglected the relative velocity of the cold relics, that could be estimated by considering the equipartition theorem:
\begin{equation}
    \frac{3}{2}T = \frac{1}{2}mv^2 \;,
\end{equation}
 from which we get:
\begin{equation}
    v = \left(\frac{3}{x_{\rm freeze-out}}\right)^{\frac{1}{2}} \approx 0.3\;.
\end{equation}
Now let us consider the cross section of a WIMP:
\begin{equation}
    \sigma_{\rm EW} \simeq G_{F}^2 T_{\rm freeze-out}^2 \approx G_{F}^2\left(\frac{E_{EW}}{20}\right)^2 \approx 10^{-8} \mbox{ GeV}^{-2}\;, 
\end{equation}
where $E_{EW} \approx 200$ GeV is the electroweak energy scale. We have then:
\begin{equation}\label{WIMPmiracle}
    \Omega_{\chi} \approx 0.01\;x_{\rm freeze-out} \approx 0.3\;,
\end{equation}
which is of the correct order of the observed value $\Omega_{\rm DM} \approx 0.26$. This is the so-called \textit{WIMP miracle}. However, note that in general this is not a special feature of WIMPs. Indeed, in order to guarantee the result in Eq.~\eqref{WIMPmiracle} one has just to use the following values:
\begin{equation}
	m_{\chi}\sigma M_{P} \gg 1\;, \qquad \sigma \approx 10^{-8}\;{\rm GeV}^{-2}\;,
\end{equation}
from which we can read the pretty weak constraint:
\begin{equation}
    m_{\chi} \gg 0.1\, \mbox{eV}\;,
\end{equation}
Now it is natural to ask ourselves which is the range of masses expected for cold relics. Since we want to preserve the unitarity of the interaction (see e.g. \cite{Weinberg:1995mt} about the upper bound on a cross section coming from the requirement of unitarity of the scattering matrix), the cross section is bounded as follows:
\begin{equation}
    \sigma \lesssim \frac{4\pi}{m_{\chi}^2}\;,
\end{equation}
so that
\begin{equation}
    \frac{\Omega_{\chi}}{0.2} \gtrsim 10^{-8}\;{\rm GeV}^{-2}\frac{m_{\chi}^2}{4\pi}\;,
\end{equation}
from which we finally get the unitarity limit:\footnote{Note that this bound is not valid in presence of Sommerfeld enhancements annihilation, see for example Ref. \cite{Feng:2010zp}.}
\begin{equation}
    \left(\frac{m_{\chi}}{120 \, {\rm TeV}}\right)^2 \lesssim 1\;.
\end{equation}
On the other hand, for WIMPs the cross section is given by $\sigma \sim G_{F}^2 m_{\chi}^2$. Therefore, the smaller the WIMP mass is the earlier the freeze-out takes place and the higher the today WIMP abundance would be, eventually even overclosing the universe. The bound for which $\Omega_\chi = 1$ is called the \textit{Lee-Weinberg} bound \cite{Lee:1977ua}. Plugging instead $\sigma \sim G_{F}^2 m_{\chi}^2$ into Eq.~\eqref{Omegachieq}, we obtain:
\begin{equation}
\Omega_{\chi} h^2 \approx 0.1 \frac{10^{-8}\;{\rm GeV}^{-2}}{G_{F}^2m_{\chi}^2} \approx 0.1\left(\frac{10 \,{\rm GeV}}{m_{\chi}} \right)^2\;,  
\end{equation}
showing that a WIMP particle with mass $m_\chi = 10-100$ GeV provides the correct present DM abundance.


\subsection{Boltzmann equation}

So far we have used qualitative arguments in to characterise a relic density, but of course precise calculations are based on the solution of the Boltzmann equation, which can be written in the following schematic way:
\begin{equation}
    \hat L\left[ f \right] = \hat C\left[f \right]\;,
\end{equation}
where $\hat L$ is the Liouville operator and $\hat C$ the collision term (also an operator). The former is given by:
\begin{eqnarray}
	\hat L_{\rm NR} = \frac{d}{dt} + \frac{d\textbf{x}}{dt}\cdot{\nabla_{x}} + \frac{d\textbf{v}}{dt}\cdot{\nabla_{v}}\;,\\
    \hat L_{\rm cov} = p^{\alpha}\frac{\partial}{\partial x^{\alpha}} - \Gamma^{\alpha}_{\beta \gamma} p^{\beta} p^{\gamma}\frac{\partial}{\partial p^{\alpha}}\;,
\end{eqnarray}
in its non-relativistic and covariant forms (boldface letters represent usual 3-vectors and the dot a scalar product). The latter simplifies in a FLRW background because in this case the probability density function $f$ depends only on the energy and on the time only due to the cosmological principle of homogeneity and isotropy:
\begin{equation}
    f\left(\textbf{x}, \textbf{p},t\right) \rightarrow f\left(E,t\right)\;,
\end{equation}
and the covariant Liouville operator can be written in the following form:
\begin{equation}
    \hat L\left[f\right] = E\frac{\partial f}{\partial t} - H|\textbf{p}|^2\frac{\partial f}{\partial E}\;.
\end{equation}
The number density of a given species is given by the integral over the reduced phase space of $f$, i.e. the integral over the proper momentum volume:
\begin{equation}
    n(t) = g\int \frac{d^3 p}{\left(2\pi\right)^3} f\left(E,t\right)\;,
\end{equation}
where $g$ is, as usual, the number of spin (or helicities, for a massless species) states. Using this definition into the expression for the Liouville operator integrated in the momentum space we get for the left hand side of the (integrated) Boltzmann equation:
\begin{equation}
    g\int \frac{d^3 p}{\left(2\pi \right)^3}\hat L\left[f\right] = \frac{d n}{d t} + 3Hn\;.
\end{equation}
Assuming a 2-to-2 interaction among species of the type 
\begin{equation}
	1 + 2 \leftrightarrow 3 + 4\;,
\end{equation}
the collision factor integrated in the momentum space of particle 1 can be written in the following form (see e.g. Ref.~\cite{Kolb:1990vq}):
\begin{equation}
    g_1\int \frac{d^3 p_1}{\left(2\pi \right)^3} \hat C\left[f_1,f_2,f_3,f_4\right] = -\langle\sigma v_{\rm M\o l}\rangle\left(n_1 n_2 - n_{1}^{\rm eq}n_{2}^{\rm eq}\right)\;.
\end{equation}
In the above formula, the superscripts ``eq'' refers to quantities (in this case the number densities) in thermal equilibrium, i.e. the expressions in Eqs.~\eqref{nrel} and \eqref{nnonrel} that we have already encountered. The average $\langle\cdots\rangle$ is a thermal one, to be specified in a moment, $\sigma$ is the total cross section of the interaction between species $1$ and $2$:
\begin{equation}
    \sigma = \sum_{f} \sigma_{12} \rightarrow f\;,
\end{equation}
i.e. summed over all the possible final states, and where we have defined:
\begin{equation}
v_{\rm M\o l} \equiv \frac{\sqrt{\left(p_1\cdot p_2 \right)^2 -m_{1}^{2}m_{2}^{2}}}{E_1 E_2}\;,   
\end{equation}
where $p_1\cdot p_2$ is the product of the two four-momenta of the particles 1 and 2, $m_1$ and $m_2$ are their masses and $E_1$ and $E_2$ their energies. This $v_{\rm M\o l}$ is the so-called \textit{M\o ller velocity} and allows one to write the interaction rate in a covariant form, similar to the non-covariant expression, i.e. $\Gamma = n_1 n_2 \sigma v_{\rm M\o l} $. The thermal average is defined as:
\begin{equation}
\langle\sigma v_{M\o l}\rangle \equiv \frac{\int \sigma v_{\rm M\o l}e^{-E_1/T}e^{-E_2/T}d^3p_1 d^3p_2}{\int e^{-E_1/T}e^{-E_2/T}d^3p_1 d^3p_2}\;.
\end{equation}
If the particle 1 and 2 are the same, i.e. the interaction under investigation is an annihilation process, Boltzmann equation can be finally written as follows:
\begin{equation}
    \frac{dn}{dt} + 3Hn = \langle\sigma v\rangle\left(n^2_{\rm eq} - n^2\right)\;.
\end{equation}
This equation must be solved numerically. Typically one expands $\langle\sigma v\rangle$ in powers of the relative velocity $v$ and the zero order term, i.e. assuming a constant $\langle\sigma v\rangle$, is usually called the \textit{s-wave approximation}.

The above Boltzmann equation is the standard tool by means of which we are able to compute relic abundances. However, there exist important exceptions, such as \textit{resonances}, \textit{thresholds} and \textit{co-annihilation}, for which that standard tool fail. We do not discuss them here, but refer the interested reader to Ref.~\cite{Griest:1990kh}.


\subsection{Modified expansion history and relic abundance: the Kination example}

Until now we have considered what happens to the thermal history of a certain particle species by focusing on the left hand side of the freeze-out condition $\Gamma = n\sigma v \simeq H$. We have shown that e.g. for WIMPs the resulting relic abundance $\Omega_{\chi}$ is in good agreement with the observed value, i.e. $\rho_{\chi,0} \approx \rho_{0}$. On the other hand, it is worth to also study the implications of modifying instead the expansion history of the universe, e.g. the evolution of the Hubble factor $H$. To this extent, let us consider a \textit{quintessence} dark energy model accounting for the dynamics of $H$, e.g. an homogeneous real scalar field $\phi$, whose energy density and pressure are defined as follows:
\begin{eqnarray}
    \rho_{\phi} = \frac{1}{2}\left(\frac{d\phi}{d t} \right)^2 + V(\phi)\;,\qquad  P_{\phi} = \frac{1}{2}\left(\frac{d\phi}{d t} \right)^2 - V(\phi)\;.
\end{eqnarray}
For simplicity, we assume a constant equation of state $\omega_\phi \equiv P_{\phi}/\rho_{\phi}$, which implies, solving the continuity equation, that $\rho_{\phi} \propto a^{-3\left(1 + \omega_\phi\right)}$. When the field kinetic energy dominates over the potential one can see that $\omega_\phi \approx 1$ and $\rho_\phi \propto a^{-6}$. This phase is called \textit{Kination} \cite{Salati:2002md}.

Let us introduce a scale of temperature $T_{\rm KRE}$, which represents the temperature of kination and radiation equality, i.e. when the energy density of the scalar field and that of the radiation component are equal. We can express the Hubble factor above this threshold and the freeze-out condition as follows:
\begin{equation}
    H \approx \frac{T^2}{M_{P}}\frac{T}{T_{\rm KRE}} \approx n\sigma v\;,
\end{equation}
since for $T > T_{\rm KRE}$ the scalar field dominates and thus $H \propto a^{-3} \propto T^3$, where the last proportionality comes from thermal equilibrium. Now, suppose that e.g. WIMPs decouple from the primordial plasma during the kination epoch. How would their relic abundance be affected, with respect to the standard case that we have seen earlier? First of all, using Eqs.~\eqref{Omegachi0} and \eqref{relnchiT03}, we can establish also in the present case that:
\begin{equation}
    \Omega_{\chi}^{\rm kination} = \frac{T_{0}^3}{\rho_0}x_{\rm freeze-out}\frac{n_{\rm freeze-out}}{T^{2}_{\rm freeze-out}}\;.
\end{equation}
Then, we can estimate the ratio between the relic abundance of WIMPs (or any other cold species, since the ratio we are computing cancels out the cross section) in presence of a kination epoch and without it:
\begin{equation}
    \frac{\Omega_{\chi}^{\rm kination}}{\Omega_{\chi}^{\rm standard}} \approx \frac{T^{\rm standard}_{\rm freeze-out}}{T_{\rm KRE}} \lesssim \frac{m_{\chi}}{20}\frac{1}{T_{\rm BBN}} \approx 10^4 \frac{m_{\chi}}{100 \mbox{ GeV}}\;,
\end{equation}
where we have used the approximate result $T^{\rm standard}_{\rm freeze-out} \approx m_\chi/20$ and the bound $T_{KRE} \gtrsim T_{\rm BBN}$, since the kination epoch has to take place before the Big Bang Nucleosynthesis, otherwise it would spoil it.

The above example of kination teaches us that from a modified expansion history, different from the usual radiation-dominated epoch, we can get a huge enhancement of the cold thermal relic density.


\subsection{Dark Matter after chemical decoupling}
    
It is possible for DM in the early universe to be in kinetic equilibrium still after chemical decoupling. This happens due to elastic scattering processes between DM and e.g. Standard Model fermions:
\begin{eqnarray}
     \chi + f \leftrightarrow \chi + f   \qquad \rightarrow \qquad  \Gamma = n_{\rm rel} \sigma_{\chi + f \leftrightarrow \chi + f} \;.
\end{eqnarray}
Here the relative velocity is unity because the temperature of the thermal bath is considered to be much higher than the fermion mass, which is therefore relativistic. In the case of a prototypical WIMP we assume that: 
\begin{equation}
	\sigma_{\chi + f \leftrightarrow \chi + f} \simeq G_{F}^2T^2\;.
\end{equation}
Assuming $m_\chi \gg T$, hence nonrelativistic WIMPs, then from equipartition one has on average that:
\begin{equation}
	\frac{p^2}{2m_{\chi}} \simeq T\;, \qquad p \simeq \sqrt{m_\chi T}\;,
\end{equation} 
and since $m_\chi$ is the dominant energy scale, collisions entail a momentum transfer of the order of the thermal bath temperature, i.e. $\delta p \approx T$. This implies that $(\delta p)/p \simeq \sqrt{T/m_\chi} \ll 1$. Upon a collision, a WIMP can gain or lose momentum, thus, as for the case of a random walk, after $N$ collisions a WIMP particle momentum changes on average of the relative amount $\sqrt{N}(\delta p)/p$. When this quantity is of order one, the WIMP has changed momentum significantly and this corresponds to the following number of collisions:
\begin{equation}
    N_{\rm coll} = \left(\frac{p}{\delta p}\right)^2 \simeq \frac{m_{\chi}}{T}\;,
\end{equation}
from which we can estimate a typical kinetic decoupling temperature for a WIMP particle, through the usual formula $\Gamma \approx H$:\footnote{Note that when dealing with annihilation this formula provides the freeze-out temperature, whereas when dealing with elastic scattering this formula provides the kinetic decoupling temperature.}
\begin{equation}
n_{\rm rel}\cdot\sigma_{\chi + f \leftrightarrow \chi + f}N_{\rm coll} \simeq T^3\cdot G_{F}^2T^2\cdot\frac{T}{m_{\chi}} \approx H \approx \frac{T^2}{M_P}\;,
\end{equation}
from which we obtain for the kinetic decoupling temperature $T_{\rm kd}$:
\begin{equation}
T_{\rm kd} \simeq \left(\frac{m_{\chi}}{M_P G^{2}_F}\right)^{1/4} \approx 30 \mbox{ MeV}\;\left(\frac{m_{\chi}}{100 \mbox{ GeV}} \right)^{1/4}\;.
\end{equation}
We can see how this results affects structure formation. In fact, for temperatures larger than $T_{\rm kd}$ WIMPs cannot form halos because they are coupled to the primordial plasma and thus any fluctuation in their density is washed away. We can associate to $T_{\rm kd}$ the following mass scale:
\begin{equation}
    M_{\rm cutoff} \approx \frac{4\pi}{3}\left[\frac{1}{H\left(T_{\rm kd}\right)} \right]^3 \rho_\chi\left(T_{\rm kd}\right) \approx 30\;M_{\rm Earth}\left(\frac{10 \mbox{ MeV}}{T_{\rm kd}}\right)^3\;,
\end{equation}
where Earth's mass is:
\begin{equation}
	M_{\rm Earth} \approx 3 \times 10^{-6}M_\odot\;.
\end{equation}
The above cutoff mass is the total mass of WIMPs contained inside the Hubble radius at the epoch in the history of the universe where the temperature of the thermal bath was precisely $T_{\rm kd}$.

So the first structures to collapse might be these minihalos with mass proportional to Earth's mass, which subsequently merge to form bigger and bigger halos, in a so-called \textit{bottom-up} scenario of structure formation. However, note that the kinetic decoupling scale differs significantly for different theories of DM.

The picture changes dramatically in the case of hot relics. Taking neutrinos as an example, they decouple, by definition, when $T \gg m_{\nu}$, and they can collapse and form structures only on large scales because on smaller ones they stream away (free-streaming), washing out any inhomogeneities. We can quantify this claim by computing the distance $d_{\nu}$ travelled by the neutrino up to the moment in which $T \approx m_{\nu}$, i.e. when it becomes non-relativistic, i.e. no more hot:
\begin{equation}
	d_{\nu} \approx \frac{1}{H\left(T \approx m_{\nu} \right)}\;.
\end{equation}
Assuming again $H \approx T^2/M_P$, we then have $d_{\nu} \approx M_{P}/m^{2}_{\nu}$ and thus the cutoff mass for hot relics becomes:
\begin{equation}
    M^{\rm hot\;relic}_{\rm cutoff} \approx \rho_{\nu}\left(T \approx m_{\nu}\right)d_\nu^3 \approx \frac{M^{3}_{P}}{m_{\nu}^{2}} \approx 10^{12} M_{\odot}\left(\frac{m_{\nu}}{1 \mbox{ keV}} \right)^{-2}\;,
\end{equation}
where one uses $\rho_{\nu}\left(T \approx m_{\nu}\right) = n_{\nu}\left(T \approx m_{\nu}\right)m_\nu = m_\nu^4$.

How can we compare these results with observation? Observational constraints from Lyman-$\alpha$ experiments give a cutoff mass scale of order:
\begin{equation}
M_{\rm cutoff} \ll M_{Ly-\alpha} \simeq 10^{10} M_{\odot}\;,
\end{equation}
and we can conclude that DM can be thermally produced at most at 10 keV scale. In summary, structure formation differs strikingly between hot and cold DM, the former implying a \textit{Top-Down} scenario, whereas the latter a \textit{Bottom-Up} one. In Ref.~\cite{Davis:1985rj} it has been showed that numerical simulations of structure formation in a universe with cold DM match the observed large scale structure of the universe better than the hot DM scenario, which is thus disfavoured.    


\section{Detectability of particle Dark Matter}

We know that DM particles interact gravitationally with ordinary matter, but it is important to understand if they also couple through other known or unknown interactions. This is crucial in order to establish if there is a chance to directly detect DM particles via experiments in colliders or detectors in orbit.


\subsection{Direct detection}

Detecting particles which interact weakly interaction has always been known to be difficult. For example, Bethe and Peierls in 1934 \cite{Bethe:1934qn} estimated the cross section for the weak process: 
\begin{equation}\label{neutrinoinelasticscattering}
	\bar{\nu}_e + p \rightarrow e^{+} +  n\;,
\end{equation}
to be of order: 
\begin{equation}
	\sigma_{\bar{\nu}_e + p \rightarrow e^{+} +  n} \simeq 10^{-43} \left(\frac{E_{\nu}}{\mbox{MeV}}\right)^2 \mbox{ cm}^2\;, 
\end{equation}
and concluding that \textit{it is therefore absolutely impossible to observe process of this kind}. Anyway, Bethe and Peierls position was perhaps too pessimistic, and since 1953 neutrinos have been abundantly detected in inelastic process such as the one in Eq.~\eqref{neutrinoinelasticscattering}, whereas elastic neutrino scattering was observed in 1973 in the bubble chamber \textit{Gargamelle} \cite{Hasert:1973ff}.

Let us use WIMPs again as prototypical DM particles and look for the relevant range of energies and masses that we could expect on the basis of direct detection. A mass $m_\chi$ WIMP with velocity $v$ colliding and bouncing off a much heavier nucleus of mass $m_N$ transmits a maximal recoil momentum $p = 2m_{\chi}v$. Hence the nucleus acquires an energy: 
\begin{equation}
	E_{\rm max} = \frac{(2m_\chi v)^2}{2m_N}\;.
\end{equation}
We can bound the maximal velocity that a DM particle could have with the escape velocity of the galaxy in which it is located, which is usually of order  $v_{\rm max} \approx 500$ km/s. This implies an energy of order of the keV for DM particles with mass of the order of the GeV.

We also have bounds coming from the event rate $R = K \phi \sigma$, where
\begin{equation}
	K \approx \frac{6.0\times 10^{26}}{A}\;,
\end{equation}
is the number of targets per kg of material with atomic number $A$ and the incident flux $\phi$ is:
\begin{equation}
	\phi =\frac{v\rho_{\rm DM}}{m_{\chi}}\;,
\end{equation}
from which we get:
\begin{equation}
R = \frac{0.06 \mbox{ events}}{\mbox{kg day}}\left(\frac{100}{A}\right)\left(\frac{\sigma}{10^{-38} \mbox{ cm}^2}\right)\left(\frac{\rho_{\rm DM}}{0.3 \mbox{ GeV/cm}^3}\right)\left(\frac{v}{200 \mbox{ km/s}}\right).
\end{equation}
In order to achieve direct detection experimentalists have to deal with two crucial points: enough signal events and enough background suppression, which is not an easy task considering the weakness of the DM signal. In particular, sources of noise are:
\begin{itemize}
	\item Slowly decaying ``primeval'' nuclides such as Uranium, Thorium and Potassium-40, whose abundance is about $10^{-4}$ and half-life is $\approx 10^9$ years;
	\item Rare, fast decaying trace elements like tritium and Carbon-14, whose abundance is about $10^{-8}$ and half-lives of the order of 10 years.
\end{itemize}
In order to achieve shielding from these environmental noises experiments are often conducted underground, as in the Italian \textit{Gran Sasso National Laboratory}, where 1400 tons of rock protect the experimental setup from cosmic rays. 

There is also the possibility of detecting a DM signal in the radioactive background of our planet, for example through \textit{seasonal and diurnal modulations}, see for example Refs.~\cite{Drukier:1986tm, Collar:1993ss} or by \textit{directional information}, see e.g. Ref. \cite{Spergel:1987kx}. 

Let us consider the direct detection event rate:
\begin{equation}
	\frac{dR}{dE_R} = N_Tn_\chi\langle v\frac{d\sigma}{dE_R}\rangle\;,
\end{equation}
where $N_T$ is the number of target nuclei, $n_\chi$ the DM particle density, $v$ the DM particle velocity over which the average is performed and $E_R$ is the recoil energy, which has the following expression:
\begin{equation}
	E_R = \frac{q^2}{2m_T} = \frac{\mu^2}{m_T}v^2(1 - \cos\theta)\;,
\end{equation}
where $q$ is the transferred momentum, $m_T$ is the target mass, and $\mu$ is the reduced mass. Taking the differential, one has
\begin{equation}
	dE_R = -d(\cos\theta)(\mu^2/m_T)v^2\;,
\end{equation}
and thus we can write:
\begin{equation}
\frac{dR}{dE_{R}} = -N_T\frac{\rho_{\rm DM}m_T}{m_{\chi}\mu^2}\int_{v_{\rm min}}^{v_{\rm esc}} d^3v \frac{f\left(v\right)}{v}\frac{d\sigma}{d \cos{\theta}}\;,
\end{equation}
where $f(v)$ is the velocity probability distribution function of the DM halo.

Of course, we need to calculate the scattering cross section in order to evaluate the above quantity. In the nonrelativistic limit, the scattering matrix element is given by the Fourier transform of the WIMP-nucleus potential:
\begin{equation}
\mathcal{M}(q^2) = \int d^3\textbf{r}\langle f|V(\textbf{r})|i\rangle e^{i\textbf{q}\cdot\textbf{r}}\;,
\end{equation}
where $\langle f|$ and $|i\rangle$ are the initial and final quantum vector states. To lowest order in $v$, the potential can be approximated to be composed by axial and spin-independent contact interactions:
\begin{equation}
V\left(\textbf{r}\right) = \sum_{\rm nucleons\;n} \left(G^{n}_{s} + G^{n}_{a}\textbf{s}_{\chi}\cdot\textbf{s}_{n}\right)\delta\left(\textbf{r} - \textbf{r}_n\right)\;,    
\end{equation}
where the $G_{s,a}$ represent, respectively, the effective scalar and axial couplings between DM and nucleons, and $\textbf{s}_{\chi}$ and $\textbf{s}_{n}$ are the DM particle and nucleon spins, respectively.


Assuming that a given microscopic theory of DM (i.e. some DM Lagrangian) is given, how we can infer something about the DM-nucleus cross section? The proper way to deal with this kind of questions is via \textit{Effective Field Theory} (EFT), i.e. assuming the mass of the mediator particle of the interaction between DM and Standard Model particles to be larger than the typical momentum transfer. As an example, in the case of milli-electric charged DM the cross section is of the form:
\begin{equation}
	\sigma = \frac{16\pi\alpha^2\epsilon^2Z^2\mu^2}{q^4}\;,
\end{equation}
where $\alpha$ is the fine-structure constant, $Z$ is the nucleus atomic number and $\epsilon \ll 1$ is the DM electric charge. 


\subsection{Indirect detection}

Another interesting possibility of detection is the \textit{indirect} one, namely to use \textit{debris} (i.e. Standard Model particles) of DM pair annihilation, which are likely to be relevant if DM is a cold thermal relic, or to use decay of DM particles into Standard Model (hence in principle detectable) ones. In both cases the rates are:
\begin{eqnarray}
	\Gamma_{\rm SM,\; ann} \simeq \left( \int_{V} \frac{\rho_{\rm DM}^2}{m_{\chi}^2} \right)\times\left(\sigma v\right)\times\left(N_{\rm SM,\;ann}\right)\;,\\
\Gamma_{\rm SM,\; dec} \simeq \left( \int_{V} \frac{\rho_{\rm DM}}{m_{\chi}} \right)\times\left(\frac{1}{\tau_{\rm dec}}\right)\times\left(N_{\rm SM,\; dec}\right)\;,
\end{eqnarray}
where $\rho_{\rm DM}$ is of course the DM density for a DM particle of mass $m_\chi$ ($\rho_{\rm DM}/m_\chi$ is thus the number density, for a cold DM species), $\sigma$ is the annihilation cross section and $v$ the relative velocity of the annihilating DM particles, $\tau_{\rm dec}$ is the lifetime of the unstable, decaying DM particle and $N_{\rm SM}$ are the numbers of Standard Model species (hence the subscript SM) into which DM particles can annihilate or decay.

We can infer informations about the product $\sigma v$ by DM thermal production history, but how can we obtain informations about the decay rate? If we suppose that such decay is mediated by some mass-scale $M$ particle, then a combination of the type:
\begin{equation}
	\Gamma_5 \simeq \frac{1}{M^2}m_\chi^3\;,
\end{equation}
would give:
\begin{equation}
	\tau_5 \approx 1\mbox{ s }\left(\frac{1\mbox{ TeV}}{m_\chi}\right)^3\left(\frac{M}{10^{16}\mbox{ GeV}}\right)^2\;,
\end{equation}
which is a too short-lived candidate. If we try another combination of powers, i.e.
\begin{equation}
	\Gamma_6 \simeq \frac{1}{M^4}m_\chi^5\;,
\end{equation}
then we would get
\begin{equation}
	\tau_6 \approx 10^{27}\mbox{ s }\left(\frac{1\mbox{ TeV}}{m_\chi}\right)^5\left(\frac{M}{10^{16}\mbox{ GeV}}\right)^4\;,
\end{equation}
which is more interesting, because the time factor $10^{27}$ s is much larger than the age of the universe ($10^{17}$ s). As mass scale $M$, the Grand Unification Scale $10^{16}$ GeV has been chosen above.

If we ask ourselves what we can say about the annihilation final state, our answer is highly model-dependent and different internal symmetries give very different results. For example:
\begin{itemize}
	\item If DM particles belong to an SU(2) multiplet, then we expect well-defined combinations of $\bar{Z}Z$, $\bar WW$ final states;
	\item In Universal Extra Dimension (UED) theories \cite{Hooper:2007qk}, DM is the Kaluza-Klein KK-1 mode of hypercharged gauge boson, thus the scattering matrix element $\mathcal M$ is proportional to the fermion hypercharge $Y_f$, i.e. $|\mathcal M|^2 \propto |Y_f|^4$, and the preferred annihilation modes are the up quarks $Y_{u_L} = 4/3$ and charged leptons $Y_{e_R} = 2$.
	\item We expect a special selection rule, e.g. helicity suppression for Majorana fermion (analogous to charged pion decay):
	\begin{equation}
		|\mathcal M|^2 \propto m_f^2\;.
	\end{equation}
\end{itemize}
The decay/annihilation rate of DM can be constrained indirectly in several ways, a possible classification being the following:
\begin{itemize}
\item Very indirect. Looking for DM effects induced in astrophysical objects or in cosmological observations.
\item Pretty indirect. Using probes that do not trace back to the annihilation event, since their trajectories are bent as the particles propagate. For example, cosmic rays.
\item Not-so-indirect. Using neutrinos and gamma rays, which have the great added advantage of traveling in straight lines.
\end{itemize}


\subsubsection{Very indirect probes}

Examples of very indirect probes include e.g.
\begin{itemize}
\item Solar Physics. It is possible that DM could affect Sun's core temperature, or the sound speed in its interior;
\item Neutron Star Capture. DM can lead to Neutron star capture that eventually leads to the formation of black holes (notably e.g. in the context of asymmetric DM);
\item Supernova and Stars, in which DM could be responsible for cooling processes; 
\item Protostars, e.g. WIMP-fueled population-III stars;
\item Planets warming;
\item Cosmological observation, where the DM content has strong implication for e.g. Big Bang Nucleosynthesis, or the Cosmic Microwave Background spectrum, also affecting the time of the recombination, or the structure formation process.
\end{itemize}


\subsubsection{Pretty indirect probes} 

A good idea is to use rare cosmic rays events where antiprotons and positrons are relatively abundant due to inelastic scattering off interstellar medium protons.

An interesting probe is given by antideuterons, or even anti-$^3$He, for which the relevant process is:
\begin{equation}
	p + p \rightarrow p + p + \bar p + p + \bar n + n\;,
\end{equation}
and which has a large energy threshold, about 17 GeV, so typically large momentum, even if caused by  very low momentum DM particles.

Positrons, and in part antiprotons, have attracted attention due to ``anomalies'' reported by the detector PAMELA (part of the AMS-02 project mounted on the International Space Station) \cite{Adriani:2011cu}.

The general scheme for treating galactic CR is with diffusion (the so-called leaky-box) models; defining the differential in the energy number density of cosmic rays as $\frac{dn}{dE} = \psi\left(\textbf{x}, E, t \right)$ we can write a master equation of the form:
\begin{eqnarray}
	\frac{\partial}{\partial t}\psi = D\left(E\right)\nabla^2\psi + \frac{\partial}{\partial E}\left(b\left(E\right)\psi\right) + Q\left(\textbf{x}, E, t \right)\;,
\end{eqnarray}
where $D(E)$ represents the diffusion coefficient, $b(E)$ describes energy losses and $Q$ include all the possible sources. Note that the above master equation can be made arbitrarily more sophisticated, for example adding the effect of cosmic rays convection, or diffusive re-acceleration, or considering fragmentations and decay. However, we do not consider these possibilities here.
 
Let us assume assume as boundary conditions a cylindrical slab for our galaxy with typical dimensions of $R \approx 10$ kpc for the radius and $h \approx 1$ kpc for the height. Let us also assume an energy dependence for the diffusion coefficient of the following form:
\begin{equation}
	D\left(E\right) \approx D_0 \left(\frac{E}{E_0}\right)^{\delta}\;,
\end{equation}
with $E_0 \approx 1$ GeV, $D_0 \approx 10^{28}$ cm$^2$/s and $\delta \approx 0.7$. We can furthermore simplify the diffusion equation assuming a steady-state regime, i.e. assuming $\psi$ to be constant. In this case the typical diffusion and energy loss time-scales look like: 
\begin{equation}
	\tau_{\rm diff} \approx \frac{R^2}{D_0}E^{-\delta}\;, \qquad \tau_{\rm loss} \approx \frac{E}{b\left(E\right)}\;,
\end{equation}
and the diffusion equation simplifies as follows:
\begin{equation}
-\frac{\psi}{\tau_{\rm diff}} -\frac{\psi}{\tau_{\rm loss}} + Q = 0\;,    
\end{equation}
and has the simple solution: 
\begin{equation}
    \psi \approx Q\frac{\tau_{\rm diff}\tau_{\rm loss}}{\tau_{\rm diff} +\tau_{\rm loss}}\;.
\end{equation}
We should distinguish distinguish among different energy phases. For cosmic rays accelerated via a Fermi mechanism, i.e. through multiple reflections in a magnetic field and with no energy losses (so in this case we neglect $\tau_{\rm loss}$), then one has: 
\begin{equation}
	Q \sim E^{-2} \quad \Rightarrow \quad \psi \approx E^{-2 -\delta} \approx E^{-2.7}\;,
\end{equation}
which is a good estimate when considering cosmic rays protons, for which indeed energy losses are irrelevant. In the case of cosmic rays electrons instead energy losses are efficient only above a certain energy threshold, so that we can write:
\begin{equation}
    b_{e}(E) \simeq b^{0}_{\rm IC}\left(\frac{u_{\rm ph}}{\mbox{eV/cm}^3}\right)\cdot E^2 + b^{0}_{\rm sync}\left(\frac{B}{1 \mu\mbox{G}}\right)^2\cdot E^2\;,
\end{equation}
where $u_{\rm ph}$ is the background photon energy density, $B$ is the ambient magnetic field (measured in micro-Gauss $\mu$G), $b^{0}_{\rm IC} \approx 0.76$ and $b^{0}_{\rm sync} \approx 0.025\times 10^{-16}$ GeV/s represent the energy loss coefficients for inverse Compton scattering and synchrotron radiation emission, respectively. Therefore, we expect a broken power-law for $\psi(E)$, since:
\begin{eqnarray}
\label{primarylowE}	\psi_{\rm primary,\;low-energy} \approx Q\cdot\tau_{\rm diff} \approx E^{-2 - \delta} \approx E^{-2.7}\;,\\
 \label{primaryhighE}   \psi_{\rm primary,\;high-energy} \approx Q\cdot\tau_{\rm loss} \approx E^{-2} \cdot \frac{E}{E^2} \approx E^{-3}\;.
\end{eqnarray}
Instead, the secondary-to-primary (secondary cosmic rays are those produced from the original particles which are accelerated to high energies and constitute the primary ones) ratios are generically:
\begin{equation}
	\frac{\psi_{e^+}}{\psi_{e^-}} \approx E^{-\delta}\;.
\end{equation}
The broken power law in Eqs.~\eqref{primarylowE} and \eqref{primaryhighE} is in very good agreement with observation, but the prediction for the secondary-to-primary ratio is at odds with the observed rising positron fraction. Indeed:
    \begin{itemize}
        \item There is no excess of antiprotons, so DM should be leptophilic, which is possible but not generic;
        \item There is no observed secondary radiation due to bremsstrahlung or inverse Compton scattering;
        \item A very large pair annihilation rate is required for thermal production, which leads to unseen gamma-ray or radio emission, i.e.
       \begin{equation}
       	\langle\sigma v\rangle \approx 10^{-24}\frac{\rm cm^3}{\rm s}\cdot\left(\frac{m_{\chi}}{100\;\rm GeV}\right)^{1.5}\;.
       \end{equation}
    \end{itemize}
A possible way out is to consider the presence of a nearby point sources, such as a pulsar, that injects a burst of positrons and that entails the following correction:
\begin{equation}
        \psi \propto Q\cdot\exp\left[-\left(\frac{r}{r_{\rm diff}}\right)^2\right]\;.
\end{equation}
Then, we can estimate the age and the distance from us of the putative pulsar as follows (using $E = 100$ GeV): 
\begin{eqnarray}
	 t_{\rm pulsar} \ll  \tau_{\rm loss} = \frac{E}{b\left(E\right)} \approx 3 \mbox{ Myr}\;,\\
    {\rm distance} \ll \sqrt{D(E)t_{\rm pulsar}} \approx 3 \mbox{ kpc}\;.
\end{eqnarray}
An interesting possibility is to use anisotropy in their relative distribution in order to disentangle pulsars from DM contributions, but complications appear since the Larmor radius of a typical pulsar magnetic field is of the order of the solar system size, so a very high spatial resolution is needed.
    
    
\subsubsection{Not-so-indirect probes}

Even if very hard, it is possible to detect neutrinos from astrophysical sources. The idea is very interesting, since DM could be gravitationally captured in celestial bodies, here experiencing accretion and starting annihilation. If the latter process is in equilibrium, we expect that a large flux of neutrinos could escape, since they have a large mean free path in matter.

The best target is the Sun, since it is sufficiently large, very close to us and naturally emits low-energy neutrinos. Let us estimate the above-mentioned process quantitatively. First of all, Sun's capture rate $C^{\odot}$ can be written down as follows:
\begin{equation}
	C^\odot = \phi_\chi\cdot\left(\frac{M_\odot}{m_p}\right)\cdot\sigma_{\chi-p}\;, \qquad \phi_\chi \sim n_\chi\cdot v_{\rm DM} = \frac{\rho_{\rm DM}}{m_\chi}\cdot v_{\rm DM}\;,
\end{equation}
where the capture cross-section (DM particle $\chi$ captured by e.g. a proton $p$) can be estimated as:
\begin{eqnarray}
	\sigma_{\chi-p}^{\rm spin-dependent} \lesssim 10^{-39}\;{\rm cm}^2\;,\\
	\sigma_{\chi-p}^{\rm spin-independent} \lesssim 10^{-44}\;{\rm cm}^2\;,
\end{eqnarray}
and hence:
\begin{equation}
	C^\odot \approx \frac{10^{23}}{\rm s}\left(\frac{\rho_{\rm DM}}{0.3\;{\rm GeV/cm}^3}\right)\cdot\left(\frac{v_{\rm DM}}{300\;{\rm km/s}}\right)\cdot\left(\frac{100\;{\rm GeV}}{m_\chi}\right)\cdot\left(\frac{\sigma_{\chi-p}}{10^{-39}\;{\rm cm}^2}\right)\;.
\end{equation}
The number of accreted DM particles $N$ is:
\begin{equation}
	\frac{dN}{dt} = C^\odot - A^\odot[N(t)]^2 - E^\odot N(t)\;,
\end{equation}
where
\begin{equation}
	A^\odot \simeq \frac{\langle \sigma v\rangle}{V_{\rm eff}}\;, \qquad V_{\rm eff} \simeq 10^{28}\;{\rm cm}^3\;\left(\frac{m_\chi}{100\;{\rm GeV}}\right)^{3/2}\;,
\end{equation}
being $A^\odot$ the annihilation rate and $E^{\odot}$ the evaporation rate, taking into account the effect of evaporation of DM particles due to their thermal velocity. The annihilation can be written as follows:
\begin{equation}
	\Gamma_A = \frac{1}{2}A^\odot[N(t^\odot)]^2 = \frac{C^\odot}{2}\left[\tanh\left(\sqrt{C^\odot A^\odot}t^\odot\right)\right]^2\;,
\end{equation}
where $t^\odot \approx 4.5$ Gyr, or $10^{17}$ s, is the age of the Sun. Defining the equilibration time as:
\begin{equation}
	t^{\rm eq} \equiv \frac{1}{\sqrt{C^\odot A^\odot}} \ll t^\odot\;,
\end{equation}
then thermal DM is in equilibration as long as WIMP-nucleon cross section is larger than $10^{-41}$ cm$^{2}$. With equilibration, the flux of neutrinos of flavour $f$ only depends on the capture rate and is:
\begin{equation}
	\frac{dN_{\nu_f}}{dE_{\nu_f}} = \frac{C^\odot}{8\pi(D^\odot)^2}\left(\frac{dN_{\nu_f}}{dE_{\nu_f}}\right)_{\rm inj}\;,
\end{equation}
where $D^\odot$ is the diffusion coefficient of the Sun and the inj subscript on the right hand side refers to the injected flux, i.e. the intrinsic flux which the Sun naturally produces.

The number of events of $\mu$-neutrinos detection at the IceCube\footnote{\url{https://icecube.wisc.edu}.} observatory is expected to be:
\begin{equation}
N_{\rm events} = \int dE_{\nu_{\mu}}\int dy \left[A_{eff}\frac{dN_{\nu_{\mu}}}{dE_{\nu_{\mu}}}\frac{d\sigma}{dy}(E_{\nu_{\mu}}, y)R_{\nu_{\mu}}(E_{\nu_{\mu}})\right]\;.
\end{equation}
The most probable final states upon DM annihilation are $\bar WW$, $\bar ZZ$ or leptons-antileptons pairs. So far no anomalous events from the Sun have been observed, and detection from Earth is even less promising. Instead, an opportunity comes from low-energy-threshold subdetectors, such as DeepCore and PINGU (both belonging to IceCube).

Another interesting idea is to look for photons coming from a possible coupling between DM and Standard Model particles. For example, a ``smoking-gun'' signal would be a monochromatic, spatially extended gamma-ray emission, because a monochromatic gamma-ray line with a diffuse morphology has no astrophysical counterparts. From data by the FermiLAT\footnote{\url{https://fermi.gsfc.nasa.gov/science/instruments/lat.html}} Bringmann \cite{Bringmann:2012vr} and Weniger \cite{Weniger:2012tx} identified a 130 GeV line from the galactic center that was initially explained by recurring to DM annihilation. Unfortunately, the signal turned out to be a statistical fluke, since it was too narrow and its significance did not increase with time. Finally, when the data were processed with the Pass 8 software no excess was detected.

Generally, we classify photon signals into two categories; in the primary one we have prompt emission of photons and bremsstrahlung radiation;\footnote{The interested reader might try to run a simulation with PITHIA, \url{http://home.thep.lu.se/Pythia/}.} in the secondary one, photons undergo inverse Compton scattering or constitute synchrotron radiation.

The average photon energies of the second kind of processes (inverse Compton scattering or synchrotron radiation) are of the order:  
\begin{equation}
\langle E^{'}_0\rangle \approx \frac{4}{3}\gamma^2_eE_0\;,
\end{equation}
where $\langle E^{'}_0\rangle$ is the average final energy of the scattered photon, $\gamma_e$ is the electron Lorentz factor and $E_0$ is the photon thermal energy which can be, for example:
\begin{eqnarray}
    E_{0,CMB} \approx 2 \times 10^{-4} \mbox{ eV} \;, \quad E_{0,\rm starlight} \approx 1 \mbox{ eV} \;, \quad  E_{0, \rm dust} \approx 0.01 \mbox{ eV}\;,
\end{eqnarray}
for the CMB thermal bath or for different instances of interstellar medium. The typical energy scales $E_e$ for electrons produced in DM decay or annihilation and the energy of scattered CMB photons can be estimated as follows:
\begin{eqnarray}
    E_e \approx \frac{m_{\chi}}{10} \quad \Rightarrow \quad \gamma_e \approx 2 \times 10^4\left(\frac{m_{\chi}}{100 \mbox{ GeV}}\right)\;,\\
    \langle E^{'}_{0,CMB}\rangle \approx 10^5 \mbox{ eV} \left(\frac{m_{\chi}}{100 \mbox{ GeV}} \right)^2\;,
\end{eqnarray}
while the typical frequency of the synchrotron radiation is:
\begin{equation}
    \frac{\nu_{\rm sync}}{\text{MHz}} \approx 10\left(\frac{E_e}{\text{GeV}} \right)^2 \left(\frac{B}{\mu G}\right) \approx \left(\frac{\gamma_e}{1000}\right)^2\left(\frac{B}{\mu G}\right)\;. 
\end{equation}
The prompt emission of photons instead depends only on the annihilation final state and the target of choice; the flux of photons $\phi_{\gamma}$ produced by DM from a given direction $\psi$ within a solid angle $\Delta \Omega$ is:
\begin{eqnarray}
    \phi_{\gamma} = \frac{\Delta \Omega }{4 \pi } \left\{\frac{1}{\Delta \Omega}\int d\Omega \int dl\left(\psi \right)\left(\rho_{DM}\right)^2  \right\}\frac{\langle\sigma v\rangle}{2m_{\chi}^2}\sum_f \frac{dN_{\gamma}^{f}}{dE_{\gamma}}\;,
\end{eqnarray}
where $\int dl$ is an integration along the line of sight and $\sum_f$ is the sum over all the possible final states. It is useful to define the socalled $J$-factor as the expression above between curly brackets:
\begin{equation}
    J(\Delta\Omega, \psi) \equiv \frac{1}{\Delta \Omega}\int d\Omega \int dl\left(\psi \right)\left(\rho_{DM}\right)^2\;,
\end{equation}
in order to characterise typical sources of radiation for an angular region that varies from 1 to 0.1 degrees. Some of them are the following:
\paragraph{Dwarf Spheroidal Galaxies}
\begin{itemize}
    \item Draco,  $J \approx 10^{19} $ GeV$^2$/cm$^5$ $\pm$ a factor 1.5;
    \item Ursa Minor, $J \approx 10^{19}$ GeV$^2$/cm$^5$ $\pm$ a factor 1.5;
    \item Segue, $J \approx 10^{29}$ GeV$^2$/cm$^5$ $\pm$ a factor 3.
\end{itemize}
\paragraph{Local Milky-Way like galaxies}
\begin{itemize}
    \item M31, $J \approx 10^{19}$ GeV$^2$/cm$^5$.
\end{itemize}
\paragraph{Local clusters of galaxies}
\begin{itemize}
    \item Fornax, $J \approx 10^{18}$ GeV$^2$/cm$^5$;
    \item Coma, $J \approx 10^{17}$ GeV$^2$/cm$^5$;
    \item Bullet, $J \approx 10^{14}$ GeV$^2$/cm$^5$.
\end{itemize}
\paragraph{Galactic center}
\begin{itemize}
    \item $0.1^{\circ}$, $J \approx 10^{22-25}$ GeV$^2$/cm$^5$;
    \item $1^{\circ}$, $J \approx 10^{22-24}$ GeV$^2$/cm$^5$.
\end{itemize}
In order to have detection we need to collect enough photons (i.e. we need a high signal-to-noise ratio). The number of detected photons is given by:
    \begin{equation}
        N_\gamma \approx A_{\rm eff} t_{\rm obs}\phi_{\gamma} \approx 10^{-20}\frac{J}{\text{GeV}^2/\text{cm}^5} \; ,
    \end{equation}
where $A_{\rm eff}$ is the effective area of the detector and $t_{\rm obs}$ is the typical observation times, and in the second approximation we considering typical values from the Fermi-LAT experiment. The energy range is of order:
\begin{equation}
    \int dE_{\gamma}\frac{dN_{\gamma}}{dE_{\gamma}} \approx \frac{m_{\chi}}{\text{GeV}}\;,
\end{equation}
so that the photon flux can be written as:
\begin{equation}
    \phi_{\gamma} = \left(\Delta \Omega\cdot J\right)\frac{1}{8\pi} \frac{\langle\sigma v\rangle}{m_{\chi}^{2}} m_{\chi} \approx 10^{-32}\frac{1}{\text{cm}^2 \text{s}}\left(  \frac{J}{\text{GeV}^2 / \text{cm}^5}\right) \; ,
    \end{equation}
from which we obtain a $J$ factor of order:
\begin{equation}
        J_{\rm tot} \approx 10^{20} \text{GeV}^2 /\text{cm}^5\;,
    \end{equation}
that is a bit bigger than the typical values of $J$ given above for dwarf spheroidal galaxies. Finally, we can compute the following rough constraint on DM pair annihilation rate depending on the DM mass:
    \begin{equation}
        \langle\sigma v\rangle \lesssim 3 \times 10^{-26} \frac{\text{cm}^3}{\text{s}}\left(\frac{30 \, \text{GeV}}{m_{\chi}} \right) \; .
    \end{equation}
For a more detailed analysis see for example Ref. \cite{Ackermann:2011wa}. In addition, we have also production of monochromatic photons from $\chi + \chi \rightarrow \gamma + \gamma $, for which:
    \begin{equation}
    \frac{\langle\sigma v\rangle_{\chi + \chi \rightarrow \gamma + \gamma}}{\langle\sigma v\rangle_{\rm tot}} \approx \frac{\alpha^2}{16 \pi^2}\;.
    \end{equation}
        

\subsubsection{The Galactic center excess}

In Ref.~\cite{Goodenough:2009gk} the authors realised, after analysing data from the the Fermi gamma-ray space telescope, that there exists an excess of gamma-rays over the background in the center of our Galaxy. Such excess was also subsequently reported independently and with a variety of different assumptions for the background in Refs. \cite{Daylan:2014rsa, Abazajian:2014fta, Gordon:2013vta}. If we ask ourselves what might produce such an excess, it is very easy to fit the data with a signal from DM annihilation;  indeed both the morphology and the spectrum are compatible with some annihilation processe, and also compatible with constraints from CMB and observations of dwarf spheroidal galaxies. On the other hand, trying to explain such an excess with emissions of astrophysical origin, e.g. unresolved pulsars, appears problematic for a variety of reasons.

However, it is possible that the starting assumptions of these analyses are in fact wrong, e.g. there is no excess at all and our models for the diffusion emission are simply inadequate for current data. Indeed, diffusion emission is a complicated physical process that involves numerous ingredients; cosmic rays are transported from their acceleration sources through the interstellar medium; most gamma-rays are produced by $p$-$p$ inelastic collisions via the production of neutral pions that decay and release radiation, or by bremsstrahlung from charged cosmic rays that decelerate due to interstellar hydrogen.

Why do we expect that our model of emission might be not realistic? The main reason is that the following assumptions are usually used to evaluate the quantity of radiation that we expect from the galactic center: $i)$ a bi-dimensional gas density distribution; $ii)$ a bi-dimensional model of cosmic rays propagation; $iii)$ the hypothesis of steady-state and $iv)$ a simplistic source distribution for the cosmic rays. It turns out that each of these assumptions costs a systematic effect which is of the same order as the excess itself. How can we improve the overall picture towards the next generation of diffuse gamma-ray models? We should consider a 3-dimensional model for the gas density distribution and for the propagation of cosmic rays, and the effect of transient processes, such as the presence of cosmic ray bursts, together with a physically motivated choice for the cosmic rays source distributions.       

In Ref. \cite{Carlson:2014cwa} the authors show how using physically-motivated models for the location of the acceleration sites of cosmic rays leads to a significant reduction of the Galactic center excess, as well as to drastic changes to its spectrum and morphology. Additionally, these novel diffuse emission models provide globally a better fit to the Galactic emission as a whole.

While significant progress is made in the direction of better and more physically motivated models for the diffuse emission, the discrimination between unresolved point sources and the diffuse emission is also part of the issue of extracting an excess and understanding its properties. Beyond and because of these considerations, we remain skeptic about firmly establishing a conclusive DM detection signal from the Galactic Center; while we still are optimistic about the possibility of detecting DM from gamma-rays.


\subsubsection{Collider production of Dark Matter particles}

In principle DM particles can be produced in collider experiments, but the expected production rate is very low if compared to Galactic DM fluxes. A possible idea is to look for anomalous events with missing energy and Standard Model particles, such as for example monojets or monophotons. We can use two different approach:
\begin{enumerate}
\item {\bf Top-down:}  pick a model and scan the parameter space (e.g supersymmetry or unified theories).
\item{\bf Bottom-up:}  use some effective field theory (EFT) or simplified models to sketch of how DM could manifest itself at colliders.
\end{enumerate}
In the EFT approach one comes up with a list of effective operators compatible with the constraints on DM coming from collider experiments \cite{Goodman:2010ku} and then apply the following algorithm: compute the production cross section and simulate the events, and finally set limits on the effective operator scale.

The main issue of such approach is that EFT is meaningful only in a certain range of energy, whose scale say $\Lambda$ is given by:
\begin{equation}
    \Lambda \simeq \frac{M}{\sqrt{g_1g_2}}\;,
\end{equation}
where $M$ is the interaction particle mediator mass, and the $g_i$'s are the couplings to Standard Model particles and particles beyond the Standard Model. This entails the question of whether the constraints that we assumed make sense, i.e. if the typical energy of the reaction (say the momentum transfer $q$) is smaller then $4\pi \Lambda$. 

An alternative approach is to use simplified models for the interactions of DM with Standard Model particles; for the coupling with a scalar $S$ and a vector $V^\mu$ mediator, we have for example the following Lagrangians:
\begin{eqnarray}
    \mathcal{L}_S \supset -\frac{1}{2}M^{2}S^2 - y_{\chi}S\chi\bar{\chi} - y^{ij}_{q}S\bar{q}_i q_j + \, \text{ h.c.} \; ,\\
    \mathcal{L}_{V} \supset -\frac{1}{2}M^2V_{\mu} V^{\mu} - g_{\chi}V_{\mu}\bar{\chi}\gamma^{\mu}\chi - g^{ij}_q V_{\mu} \bar{q}_i \gamma^{\mu}q_j + \, \text{h.c.}\;,
\end{eqnarray}
where the $y$ and the $g$'s are coupling constants, the $\gamma^\mu$ are Dirac's matrices, the symbols $\supset$ means that the contributions on the right hand side are only a part of the total Lagrangian (there still is the Standard Model one) and h.c. means ``hermitian conjugate''.

The idea is to set meaningful constraints on combinations of mediator mass and couplings constants for given DM masses. Then, we can compare with direct detection results.

An additional probe of DM within colliders is given by the invisible decay of the Higgs boson into DM, with an interaction term of the form:
\begin{equation}
    \lambda_{H \chi \bar{\chi}} \bar{\chi}\chi |H|^2 \; , 
\end{equation}
where $H$ is the Higgs scalar (not to be confused with the Hubble factor) and $\lambda_{H \chi \bar{\chi}}$ is the DM-Higgs coupling. 

The invisible Higgs decay modes may stem from decays of the Higgs boson directly into DM particles, or into new particles that in turn decay into DM particle, see for example Ref. \cite{Aad:2015txa}.


\subsubsection{Axions and Axions searches}

Axions are hypothetical particles postulated in order to solve the strong CP problem in Quantum Chromodynamics (QCD) \cite{tHooft:1976snw}. Several production mechanisms can lead to abundant production of axions in the early universe. If the universe is indeed filled with such primordial particles they can be, in principle, good DM candidates. The QCD Lagrangian is given by:
\begin{equation}
    \mathcal{L}_{QCD} = -\frac{1}{4}G^{a}_{\mu \nu} G^{a \, \mu \nu} + \sum_{j=1}^{n} \left[\bar{q}_j \gamma^{\mu}iD_{\mu}q_j - \left( m_j q_{L \, j}^{\dagger} q_{R \, j} + \text{h.c.} \right)   \right] + \frac{\theta}{32 \pi^2}\epsilon^{\mu\nu\rho\sigma}G^{a}_{\mu \nu} G^a_{\rho\sigma}\;,
\end{equation}
where $G^{a}_{\mu \nu}$ is the gauge-field (the gluon) tensor, the $n$ $q$'s are the quark fields, $D_\mu$ is the gauge-covariant derivative. The last term, where $\theta$ is a coupling and $\epsilon^{\mu\nu\rho\sigma}$ is the Levi-Civita symbol, is innocuous perturbatively, being a total derivative, but it enters phenomenologically via non-perturbative QCD effects, producing a large neutron-electron dipole moment:
\begin{equation}
    d_n \approx 5 \times 10^{-16}\theta \, e \,  \text{cm}\;,   \qquad d_n \lesssim 10^{-26} \,  e \, \text{cm}\;,
\end{equation}
where the last bound is the experimental result \cite{Afach:2015sja}. The Peccei-Quinn solution \cite{Peccei:1977hh} consists of promoting $\theta$ to a dynamical variable which is driven to vanish by its classical potential. Then one postulates a global (quasi-)symmetry of the theory $U(1)_{PQ}$ (broken by perturbative effects); such a symmetry is spontaneously broken at a scale $f_a$. In this way, the axion turns out to be the (pseudo) Nambu-Goldstone boson associated with the group $U(1)_{PQ}$. Its mass is given by:
\begin{equation}
    m_a \simeq \frac{\Lambda_{QCD}^2}{f_a} \approx 0.6 \mbox{ eV}\left(\frac{10^7 \text{GeV}}{f_a}\right)\;,
\end{equation}
where $\Lambda_{QCD} \approx 200$ MeV is the QCD energy scale. As shown in Ref. \cite{Vafa:1984xg}, the ground state of the axion potential solves the strong CP problem. These QCD effects produce effective couplings (slightly model-dependent) to fermions and photons:
\begin{equation}
\mathcal{L}_{af\bar{f}} = ig_f\frac{m_f}{f_a/N}a\bar{f}\gamma_5 f \, , \qquad \mathcal{L}_{a\gamma \gamma} = -g_{\gamma}\frac{\alpha}{\pi}\frac{a}{f_a}\textbf{E}\cdot \textbf{B} \;,
\end{equation}
where $a$ represents here the axion field, $N$ is the number of fermion species considered, $\gamma_5 = i\gamma_0\gamma_1\gamma_2\gamma_3$, $\textbf{E}$ and $\textbf{B}$ are the electric and magnetic fields and, as usual, the $g$'s are the coupling constants.

The setup is similar if we consider \textit{Axion Like Particles} (ALP): we still have a new global $U(1)$ symmetry which is spontaneously broken by a hidden Higgs-like mechanism at scale $v_h$. We parametrize the Higgs field as :
\begin{equation}
    H_h(x) = \frac{1}{\sqrt{2}}\left[v_h + h_h(x) \right]e^{ia(x)/v_h} \; .
\end{equation}
The potential for the ALP field $a(x)$ is flat, and depending on the model realizations one generates couplings to SM particles:
\begin{equation}
    \mathcal{L}_{ALP} = \frac{1}{2}\partial_{\mu}a\partial^{\mu}a - \frac{\alpha_s}{8\pi}C_{ag}\frac{a}{f_a}\epsilon^{\mu\nu\rho\sigma} G^{a}_{\mu \nu}{G}^{a}_{\rho\sigma} - \frac{\alpha}{8 \pi} C_{a\gamma}\frac{a}{f_a} \epsilon^{\mu\nu\rho\sigma}F_{\mu\nu}F_{\rho\sigma} + \frac{1}{2}\frac{C_{af}}{f_a}\partial_{\mu}a\bar{f}\gamma^{\mu}\gamma^{5}f \;,
\end{equation}
where $F_{\mu\nu}$ is the electromagnetic field tensor. Because of the coupling to Standard Model particles, in particular photons, axions decay into two photons:
\begin{equation}
    \tau_{a \rightarrow \gamma \gamma} \simeq \frac{16\pi^2}{\alpha^2}\frac{\Lambda^{4}_{QCD}}{m_a^5} \approx 10^{24} \, \text{s} \left(\frac{1 \text{ eV}}{m_a} \right)^5 \; .
\end{equation}
To have a sufficiently long-lived axion we must demand also:
\begin{equation}
    \tau \approx 10^{10} \times \left( \pi10^{7} \right) \, \text{s} \lesssim 10^{24} \text{ s} \left( \frac{1 \text{ eV}}{m_a} \right)^5 \Rightarrow m_a \lesssim 25 \text{ eV} \; , \quad f_a \gtrsim 4 \times 10^6 \text{ GeV} \; .
\end{equation}
The axions can have a dramatic impact on stars due to their Compton-like and Bremsstrahlung-like interactions:
\begin{equation}
    \gamma + e  \rightarrow a + e \; , \qquad e + Z \rightarrow a + e + Z \; , 
\end{equation}
which produce an axion luminosity, for example for the Sun, of order:
\begin{equation}
    L_a \approx  6 \times 10^{-4} \left(\frac{m_a}{1 \text{ eV}} \right)^2 L_{\odot} \; .
\end{equation}
Given the observed luminosity of the Sun, axion emission would require an enhanced nuclear energy production, which in turns would imply a larger neutrino flux.

Axions can also have impact on the physics of supernovae. If the axion mass is small enough, axions would free-stream out of supernovae, causing their cooling; the corresponding axion luminosity is of order:
\begin{eqnarray}
    L_a \approx 10^{59} \text{ ergs/s } \left(\frac{m_a}{1 \text{ eV}} \right)^2 \; .
    \end{eqnarray}
For consistency we should require that this luminosity does not exceed the value of the thermal neutrino luminosity $L_{\nu} \approx 10^{53} \text{ erg/s }$, which would happen for:
    \begin{equation}
        L_a > L_{\nu} \text{ for } m_a > 10^{-3} \text{ eV } \; . 
     \end{equation}
On the other hand, if axions are too massive, they get trapped and not contribute to supernovae luminosity efficiently. It turns out that axions would affect supernovae burst duration in the mass range:
\begin{equation}
    10^{-3} \lesssim m_a / 1 \text{ eV} \lesssim 2 \; .
\end{equation}
How can axions be produced? If they are produced thermally, this should happen via interaction among gluons and quarks of the form:
\begin{eqnarray}
    a + g \leftrightarrow \bar{q} + q \text{ or } g + g \, , \text{ or } a + q\left(\bar{q}\right)  \leftrightarrow g + q\left(\bar{q}\right) \;.
    \end{eqnarray}
This kind of process comes from the QCD Lagrangian term proportional to $\theta$, from which we can also roughly estimate the following cross section:  
\begin{eqnarray}    
\sigma_{q,g} \simeq \frac{\alpha^{3}_{s}}{\pi^2 f^2_{a}}\;, 
\end{eqnarray}
where $\alpha_s$ is the coupling constant relative to the strong interaction. From the above formula we can obtain a rough estimate of the temperature at which the processes considered above go out of equilibrium and the axions freeze-out: 
\begin{eqnarray}
    \frac{\Gamma}{H} \approx 1 \quad \Rightarrow \quad T^3 \sigma_{q,g} \approx \frac{T^2}{M_P} \; ,\\
    T_{a,\rm freeze-out} \approx 10^{11} \text{ GeV } \left(\frac{f_a}{10^{12} \text{GeV}}   \right)\; .
\end{eqnarray}
However, at lower temperatures (below the QCD phase transition scale) the interactions with pions must be considered:
\begin{eqnarray}
    \pi + \pi  \leftrightarrow \pi + a \; , \qquad \sigma_{\pi} \simeq \frac{1}{f_a^2}\;. 
    \end{eqnarray}
Such process is in equilibrium when:
    \begin{eqnarray}
    \frac{n_{\pi} \sigma_{\pi}}{H} \approx \frac{m_{\pi}M_P}{f_a^2} \gtrsim 1 \qquad \Rightarrow \qquad f_a \lesssim 5 \times 10^8 \text{GeV} \; .
    \end{eqnarray}
Finally, we obtain the density of axions as a function of their mass:
    \begin{eqnarray}
    \Omega_{a} \approx \frac{m_a}{\text{130 eV}} \; ,
\end{eqnarray}
but we know already that this does not work since hot DM is not good for structure formation. Also, other constraints on axion mass are also not good, because in order to have $\Omega_{a} \approx \Omega_{DM} \approx 0.26$ we would need $m_a \approx 10$ eV, and the resulting axion lifetime would be too short. 

But then, how about non-thermal production? We can consider for example the so-called misalignment mechanism for producing cold axions, which results in the following cosmological density, see Ref. \cite{Sikivie:2006ni}:
\begin{equation}
    \Omega_{a,\text{mis}}h^2 \approx 0.4\left(\frac{m_a}{10\;\mu\text{eV}} \right)^{-1.18}\left(\frac{\theta_1}{\pi}\right)^2 \; ,
    \end{equation}
where $\theta_1$ is a mixing angle. If we assume that the value for $\theta_1$ is given by its root mean square we get:
    \begin{eqnarray}
    \theta_{\text{RMS}} = \left(\int_{-\pi}^{\pi} d{\theta} \frac{{\theta}^2}{2\pi}\right)^{\frac{1}{2}} = \frac{\pi}{\sqrt{3}}\; ,\\
    \Omega_{a,\text{mis, RMS} } h^2 \approx 0.13 \left(\frac{m_a}{10 \;\mu\text{eV}}\right)^{-1.18} \; .
    \end{eqnarray}
On the other hand, if the distribution for $\theta_1$ has a non trivial topology, a network of axion strings and domain walls can be generated; the resulting cosmological density is in general model-dependent and the results of different groups are in general in disagreement. For example, Refs. \cite{Hiramatsu:2012gg, Sikivie:2006ni} give the following different values:  
    \begin{eqnarray}
    \Omega_{\text{strings + domain walls}}h^2 = \left(3.5 \pm 1.7 \right)\left(\frac{m_a}{10\;\mu\text{eV}}\right)^{-1.18} \; ,\\
    \Omega_{\text{strings + domain walls}}h^2 \sim 0.4 \left(\frac{m_a}{10 \;\mu\text{eV}}\right)^{-1.18} \; .
\end{eqnarray}
Axion laboratory searches include light-shining-through-wall experiments, see for example \cite{Redondo:2010dp, Ballou:2015cka} that exploit the process:
\begin{equation}
    \gamma + Ze \leftrightarrow Ze + a\;.
\end{equation}
Additional searches are conducted based on microwave cavities \cite{Sikivie:1983ip} and helioscopes \cite{Anastassopoulos:2017ftl}.


\subsubsection{Sterile neutrinos and the 3.5 keV line puzzle}

One of the few still open problems of the Standard Model of particle physics is the fact that neutrinos are not massless. A simple way out is to consider a set of $n$ gauge singlet fermions $N_a$ (the sterile neutrino field) so that:
\begin{equation}
    \mathcal{L} = \mathcal{L}_{SM} + i\bar{N}_a \slashed{\partial} N_a - y_{\alpha a} H^{\dagger} \bar{L}_\alpha N_a - \frac{M_a}{2} \bar{N}_a^c N_a + \mbox{ h.c.}\; ,
\end{equation}
where $\slashed{\partial} \equiv \gamma^\mu\partial_\mu$ and $\bar{L}_\alpha$ represent the ordinary neutrino field. The mass matrix can be computed to be the following:
\begin{equation}
    \textbf M^{\left(n + 3\right)} = \begin{bmatrix} 0 & y_{\alpha a}\langle H\rangle \\ y_{\alpha a}\langle H\rangle & \text{diag}\left(M_1, M_2 \dots M_n\right)
    \end{bmatrix} \;,
\end{equation}
where the Higgs field vacuum expectation value $\langle H\rangle$ is usually written as $iv/\sqrt{2}$ (do not confuse $v$ with any velocity). Just for simplicity, assume only one sterile neutrino family and only one ordinary neutrino family, so that the above matrix become a $2\times 2$ matrix:
\begin{equation}
    \textbf M = \begin{bmatrix} 0 & y\langle H\rangle \\ y\langle H\rangle & M
    \end{bmatrix} \;.
\end{equation}
It is easy then to compute its eigenvalues, which are:
\begin{equation}
	\lambda_\pm = \frac{M \pm \sqrt{M^2 + 4y^2\langle H\rangle^2}}{2} = \frac{M \pm \sqrt{M^2 - 2y^2v^2}}{2}\;,
\end{equation}
hence, if the condition $yv \ll M$ is satisfied, then $\lambda_+ \gg \lambda_-$. In general, since $\lambda_+\lambda_- = y^2v^2/2$, if one eigenvalue grows the other decreases and viceversa. This is the so-called \textit{Seesaw mechanism}. So if $yv \ll M$ is satisfied, we have one large eigenvalue $M$ corresponding to the sterile neutrino mass and one small $y^2v^2/M$ eigenvalue corresponding to the ordinary neutrino mass.  

Sterile neutrinos mix via explicit (but possibly very small) mixing with ordinary neutrinos; as such, they decay (into the 3 Standard Model neutrinos):
\begin{eqnarray}
    \Gamma \approx \theta^2G_F^2m_{\nu_s}^5 \approx \theta^2 \left(\frac{m_{\nu_s}}{\text{keV}}\right)^5 10^{-40} \text{ GeV} \Rightarrow \tau \approx 10^{16} \text{ s}\;,
\end{eqnarray}
where $m_{\nu_s}$ is the sterile neutrino mass, $\theta$ is the mixing angle and where we have used $\theta^{-2}\left(\frac{m_{\nu_s}}{\text{keV}}\right)^{-5} \gg 1$ because sterile neutrinos are fermions and thus $m_{\nu_s} > \text{keV}$ \cite{Tremaine:1979we}.

Sterile neutrino are produced essentially via freeze-in \cite{Hall:2009bx}. Briefly, the freeze-in mechanism relates to particles which are not in thermal equilibrium with the primordial plasma but still interact (and are produced) with some of the species which instead are in thermal equilibrium with the primordial plasma. This interaction stops being effective when the latter species freeze-out, so that the weakly interacting species are said to freeze-in. For sterile neutrinos we can estimate the production rate as follows:
\begin{equation}
    \Gamma_{\nu_s} \approx ( G_F^2 T^5)\theta^2\left(T\right)\;,
\end{equation}
where the mixing angle $\theta$ is usually considered to be constant, or in some refined models has a temperature dependence:
\begin{equation}
    \theta \rightarrow\frac{\theta}{1 + 2.4\left(\frac{T}{200 \text{MeV}}\right)^6 \left(\frac{1 \text{ keV}}{m_{\nu_s}}\right)^2} \; .
\end{equation}
Abundance of sterile neutrinos can be estimated as the following value: \cite{Dodelson:1993je}:
\begin{equation}
    \Omega_{\nu_s} h^2 \approx 0.1 \left(\frac{\theta^2}{3 \times 10^{-9}} \right)\left(\frac{m_{\nu_s}}{3 \text{ keV}}\right)^{1.8} \; .
\end{equation}
An additional important effects that can be considered are the Mikheyev-Smirnov-Wolfenstein effect \cite{Mikheev:1986gs, Wolfenstein:1977ue}, and the Shi-Fuller resonant production effect \cite{Shi:1998km}.

What if sterile neutrinos are not produced thermally? A simple alternative is to consider non-thermal production from singlet scalar coupling with the field $S$:
\begin{equation}
    \frac{h_a}{2}S \bar{N}^c_aN_a \; ,
\end{equation}
and (or) $SH^{\dagger }H$, $S^2 H^{\dagger}H$, that gives:
\begin{eqnarray}
    \Omega_{\nu_s} \approx 0.2 \left(\frac{h_a}{10^{-8}} \right)^3 \frac{\langle S\rangle}{M_S}\;.
\end{eqnarray}
Sterile neutrinos are also interesting from the standpoint of structure formation, since:
\begin{eqnarray}
    M_{\text{cutoff, hot}} \approx \left[\frac{1}{H(T = m_{\nu_s})}\right]^3\rho_{\nu_s}(T = m_{\nu_s}) \approx \left(\frac{M_P}{m_{\nu_s}^2}\right)^3 m_{\nu_s}^4 = \frac{M_P^3}{m_{\nu_s}^2} \; , \\
\frac{M_P^3}{m_{\nu_s}^2} \approx 10^{15} M_{\odot}\left(\frac{m_{\nu_s}}{30 \text{ eV}}\right)^{-2} \approx 10^{12} M_{\odot} \left(\frac{m_{\nu_s}}{1 \text{ keV}}\right)^{-2}\;.
\end{eqnarray}
Another interesting feature of sterile neutrinos is that they can provide an answer for the so called \textit{pulsar kick} problem, see for example Ref.~\cite{BisnovatyiKogan:1997an}.

What about detection of sterile neutrino DM? The key mechanism is loop-mediated sterile neutrino decay via weak interactions and mass-mixing with active neutrinos, leading to a naive estimate for the decay width and the photon flux from decay of:
\begin{eqnarray}
    \Gamma_{\nu_s \rightarrow \gamma + \nu} \approx \frac{\alpha}{16 \pi^2} \theta^2 G^2_Fm_{\nu_s}^5\;, \\
    \phi_{\gamma} = \frac{\Gamma_{\nu_s \rightarrow \gamma + \nu}}{4\pi} \frac{E_\gamma}{m_{\nu_s}} \int_{\rm fov} d\Omega \int_{\text{los}} \frac{\rho_{DM}}{m_{\nu_s}} dr(\psi) 
\end{eqnarray}
where fov means ``field of view'' and los means ``line of sight''. 

The key background radiation where to searche for sterile neutrino decay is the diffuse cosmic X-ray background (CXB):
\begin{equation}
    \phi_{CXB} \approx 9.2 \times 10^{-7} \left(\frac{E}{1 \text{ keV}}\right)^{-0.4} \text{cm}^{-2} \text{s}^{-1}\text{arcmin}^{-2} \approx 10^{-4} \text{cm}^{-2} \text{s}^{-1} \;.
    \end{equation}
This implies the following constraints on the mixing angle and mass:
\begin{eqnarray}
    \phi_{\gamma} \approx 10^{-4} \text{cm}^{-2} \text{s}^{-1}\left(\frac{\theta^2}{10^{-7}} \right)\left(\frac{m}{1 \text{ keV}}\right)^4 \left(\frac{J}{10^{18} \text{ GeV/cm}^2}\right) \; , \\
    \left(\frac{\theta^2}{10^{-7}}\right)\left(\frac{m}{1 \text{ keV}}\right)^4 \lesssim 1 \;.
\end{eqnarray} 
In summary, sterile neutrinos are:
\begin{itemize}
    \item SU(2)$_L$ gauge singlet, but have a small mixing angle with active neutrinos;
    \item Cosmologically viable candidates of DM \cite{Dodelson:1993je};  
    \item not stable because they decay via mixing with active neutrinos.
\end{itemize}
Claims have been made that sterile neutrino dark matter decay might have already been detected \cite{Bulbul:2014sua, Boyarsky:2014jta, Jeltema:2014qfa}. An emission line of 3.5 keV is roughly compatible with all the above listed properties. Unfortunately, it is also compatible with the emission line from atomic transition of highly ionised atoms, in particular, potassium  K XVIII has two lines near 3.5 keV.

So how do we tell these potassium lines apart from a sterile neutrino signal? We should try to predict K XVIII line brightness using other elemental lines, but two key complications arise:
\begin{itemize}
    \item Plasma temperature;
    \item Relative elemental abundances.
\end{itemize}
In Ref.~\cite{Bulbul:2014sua} it is argued against K XVIII since predictions for a potassium 3.5 keV line is too low (by a factor 20 for solar abundances). However, this prediction makes two mistakes based on the two complications above. In particular the authors of \cite{Bulbul:2014sua} use very large plasma temperature, which highly suppresses the potassium emission. Moreover, they underestimate the abundance of potassium of a factor $\approx 10$.

The authors of Ref.~\cite{Jeltema:2014qfa} show that for clusters of galaxies and for our galaxy K XVIII line emission could explain the 3.5 keV line. Are there are other possible tests? Of course, we can look elsewhere, or we can use some different probe rather than the spectrum. About the former option, there is no trace of signal from dSph galaxies \cite{Malyshev:2014xqa}, nor from stacked galaxies and groups with low plasma temperature \cite{Anderson:2014tza}, nor from Andromeda \cite{Jeltema:2014qfa}. Also, there is no signal from dedicated observation of Draco dSph \cite{Jeltema:2015mee}. What about using something different than spectrum? We can use morphology, e.g. look at where the 3.5 keV radiation come from. However, morphology of Perseus and the Milky way shows that decaying DM is strongly disfavoured \cite{Carlson:2014lla}.

In summary, we are left with the open question: Dark Matter or Potassium? We can look for a possible answer in the old literature: \textit{Entia non sunt multiplicanda praeter necessitatem},\footnote{William of Occam, c. 1286-1347.}. Occam's razor favours Potassium against sterile neutrinos because of the lack of signal from Draco and because of the morphology of the thermal lines. But what if it is another type of DM? Any challenge is also an opportunity and an interesting riddle for theorists; according to Redman's Theorem:\footnote{Quoted in Ref. \cite{Longair:1981jc}, Sec. 2.5.1, \textit{The psychology of astronomers and astrophysicists}.}
\begin{equation}
	\textit{Any competent theoretician can fit any given theory to any given sets of facts.}\nonumber
\end{equation}
The 3.5 keV line is an exciting excuse for a new mechanism generating a DM signal. Consider for example an interaction of the form \cite{DEramo:2016gqz}:
 \begin{equation}
     \chi_1 + f \rightarrow \chi_2 + f \longrightarrow  \chi_2 \rightarrow \chi_1 + \gamma\;,
 \end{equation}
which models interactions between DM and interstellar plasma, resulting in a signal which is proportional to the product of the DM and gas densities, and it is also a good thermal relic. Why should we be excited by this model? 
\begin{itemize}
    \item It is a new indirect detection channel;
    \item it has an unmistakable signature, free of background;
    \item Is a good model, in the sense that it is economic, with a natural UV completion and a thermal relic DM.
\end{itemize}
Moreover, it is a highly falsifiable scenario for the following reasons:
\begin{itemize}
    \item Its line shape, e.g. geometric average of thermal DM velocities, can be resolved by e.g. the Hitomi/astro-h satellite;
    \item It has unique morphology
    \item It has unique target-dependence
    \item Lines could appear anywhere in the spectrum, from eV, to UV, to X-ray.
\end{itemize}
Possible future tests include implications on structure formation and small scale structure. 


\section{Dark Matter \textit{Bestiarium}}

In this section we introduce a list of notable candidate models of DM particle.


\subsection{Gravitinos}

Gravitinos are prototypical DM with Planck-suppressed interactions \cite{Pagels:1981ke}. The prototype Lagrangian is:
\begin{equation}
\mathcal{L} =  -\frac{1}{2}\epsilon^{\mu \nu \rho \sigma}\partial_\nu\bar{\Psi}_{\mu}\gamma_5 \partial_{\rho}\Psi_{\sigma} - \frac{m_{3/2}}{4}\bar{\Psi}^{\mu}\left[\gamma_{\mu}, \gamma_{\nu} \right]\Psi^{\nu} + \frac{1}{2M_P}\bar{\Psi}_{\mu}S^{\mu}\;,
\end{equation}
and the resulting cross sections are:
\begin{eqnarray}
    \sigma\left(\tilde{G} + {G}\rightarrow \gamma + \gamma \right) = \frac{1}{576\pi}\frac{M_{\tilde{\gamma}}^2 s^2}{\left(m_{3/2}M_P\right)^4} \; ,\\
     \sigma\left(\tilde{G} + {G}\rightarrow \bar{f} + f \right) = \frac{g_f}{720\pi}\frac{ s^3}{\left(m_{3/2}M_P\right)^4} \; , 
\end{eqnarray}
where we used the symbol ${G}$ to indicate a gravitino particle. Finally, the freeze-out temperature is:
\begin{equation}
    T_{\rm freeze-out} \approx \left(\frac{m^{4}_{3/2}M_P^3}{M_{{\gamma}}^2} \right)^{1/5} \approx 450 \text{ GeV } \left( \frac{M_{3/2}}{0.1 \text{ eV}}\right)^{4/5}\left(\frac{100 \text{ GeV}}{M_{{\gamma}}} \right)^{2/5 } \;.
\end{equation}
Thermal gravitinos are hot relics (in fact they were the first super-symmetric DM candidate ever proposed) and their abundance is calculated as: 
\begin{equation}
    \Omega_{3/2} h^2 = \frac{m_{3/2}n_{3/2}T_0}{\rho_0}h^2 \approx \frac{m_{3/2}}{\text{keV}}T_{\rm freeze-out}\;.
\end{equation}
However, this calculation neglects single-gravitino processes that maintain gravitinos out of equilibrium, for example:
\begin{equation}
    V + \tilde{\lambda} \leftrightarrow V + \tilde{G} \; ; \qquad V + V \leftrightarrow \tilde{\lambda} + \tilde{G} \; ,
\end{equation} 
where $V$ is a gauge boson and $\tilde{\lambda}$ its gauginos (its supersymmetric partner). Generically, there is a gravitino overproduction problem, so we need to dilute them away:   
\begin{equation}
    \Omega_{3/2}h^2 \approx 0.00167 \frac{m_{3/2}}{1 \text{ GeV}}\frac{T_{RH}}{10^{10} \text{ GeV}}\frac{M_P^2\gamma|_{T=T_{RH}}}{T_{RH}^6} \; ,
\end{equation}
where $T_{RH}$ is the reheating temperature.

Gravitinos are also produced from next-to-lightest supersymmetric particles (NLSP) decays \cite{Bailly:2008yy}:
\begin{equation}
    \Omega_{3/2} = m_{3/2} \frac{\Omega_{NLSP}}{m_{NLSP}} \; .
\end{equation}
Light gravitinos are already ruled out, whereas heavier gravitinos are still viable for certain combinations of mass and $T_{RH}$. Moreover, they might also be NLSP with long lifetimes:
\begin{equation}
    T_{\tilde{G}} \approx \frac{M_P^2}{m_{3/2}^3} \approx 10^{14} \text{ s } \left(\frac{1 \text{ GeV}}{ m_{3/2}}\right)^3 \; .
\end{equation}
An important question is of course: can we ever hope to detect gravitinos? There are basically two strategies; the first one is to use long-lived s-leptons produced at LHC trapping them into water tanks, and then wait their decay to gravitinos. The second one is to use neutrino telescopes to search for neutrino-induced long lived s-leptons:
\begin{equation} 
    \nu + q \rightarrow \tilde{l} + \tilde{q} \; .
\end{equation}


\subsection{WIMPzillas and super-heavy Dark Matter candidates}

This interesting class of models grounds on the presence of non-thermally produced super heavy particles, the so-called WIMPzillas \cite{Kolb:1998ki}, with the following properties:
\begin{itemize}
    \item The DM particle is never in thermal equilibrium; 
    \item The particle mass is comparable to the inflaton mass say $M_{\phi}$; 
    \item The particle lifetime is much longer than the age of the universe.
\end{itemize}
 The following generic expression gives us the density of such heavy particles produced by gravitational interaction at the end of inflation \cite{Chung:2004nh, Kuzmin:1999zk}:
\begin{equation}
    \Omega_{X}h^2 = 10^{-3} \Omega_R \frac{8\pi}{3}\left(\frac{T_{RH}}{T_0}\right)\left(\frac{M_{\phi}}{M_P} \right)^2\left( \frac{M_X}{M_{\phi}} \right)^{5/2}e^{\frac{-2M_X}{M_{\phi}}} \; ,
\end{equation}
where $\Omega_R$ is the radiation density and and $T_{RH}$ is the reheating temperature. Such kind of particles could be detected in ultra high energy cosmic rays detectors.

Another interesting super heavy DM candidates are the strangelets \cite{Witten:1984rs}, i.e. macroscopic clumps of quark matter, with radius between 10 mm to 10 cm and masses around $10^9- 10^{18}$ g. They are in principle detectable via cosmic rays collisions.


\subsection{Self-interacting Dark Matter}

In order to have at least one interaction for particle during the age of the universe this class of models must satisfy:
\begin{equation}
    \frac{\Gamma}{H_0} \approx \frac{\left(\sigma_{\chi \chi} / m_{\chi}\right)\rho v}{H_0} \approx 1 \; , 
    \end{equation}
which in turn implies a cross section of order:    
    \begin{equation}
    \frac{\sigma_{\chi \chi}}{m_{\chi}} \approx 1.3 \frac{\text{cm}^2}{\text{g}} \left( \frac{1 \text{ GeV/cm}^3}{\rho} \right) \left(\frac{10 \text{ km/s}}{v} \right) \; .
\end{equation}
Events of this type could be object of study in DM halos, which can be considered as non-relativistic colliders. For example, dwarfs galaxy are equivalent to the B-Factory collider, Milky way-like galaxies reproduce events like in the LEP, while big clusters of galaxies show events that on Earth are reproducible at the LHC.

Which model of particle physics model could work for self-interacting DM? We have a quite large landscape; consider for example a kind of Dark QCD with \textit{glueballs} DM \cite{Boddy:2014yra}:
\begin{equation}
    \sigma\left( gb + gb \leftrightarrow gb + gb  \right) \approx \frac{4\pi}{\Lambda_{QCD}^2} \; ,
\end{equation}
\begin{equation}
    \rho_{gb} \approx 2(N^2 - 1 )s_0 \Lambda \left(\left.\frac{T_h}{T}\right|_{T \approx \Lambda_{QCD}}\right)^3 \; .
\end{equation}
Another possibility is given by \textit{dark atoms}, whose relic density computation is still a work in progress. Another one: \textit{light mediator} \cite{Tulin:2013teo} such that:
\begin{eqnarray}
    m_{X}v \ll m_{\phi} \; , \\
    \sigma \approx 5 \times 10^{-23} \text{ cm}^2 \left(\frac{\alpha_{X}}{0.01}\right)^2 \left(\frac{m_X}{10 \text{ GeV}} \right)^2 \left( \frac{10 \text{ MeV}}{m_{\phi}}\right)^4 \; , \\
    \sigma \sim \frac{\alpha_X^2}{m_X^2 v^4} \qquad \left(m_X v \gg m_{\phi} \right) \; .
\end{eqnarray}
Effects should turn off at large velocities but are significant at small scales or low velocities. We should assume that the mediator has a small coupling $\epsilon$ to Standard Model fermions.

Another possibility is given by a model with a complex scalar field and potential \cite{Mambrini:2015nza}:
\begin{eqnarray}
    \Phi = v + \frac{s + ia}{\sqrt{2}}  \qquad V(\Phi) = -\mu^2 |\Phi|^2 + \frac{\lambda}{4}|\Phi|^4 \; ,
    \end{eqnarray}
    which gives us the following cross section:
    \begin{eqnarray}
    \sigma \left( a + a  \leftrightarrow a + a \right) = \frac{\lambda^2 m_a^2 }{32\pi m_s^4} \left( 1 - \frac{4 m_a^2}{m_s^2} \right)^{-2} \; ,
     \end{eqnarray}
from which, using  $m_s = 1$ MeV and $\lambda \simeq 0.5$ we get 
    \begin{eqnarray}
 \frac{\sigma\left( a+a \leftrightarrow a + a \right)}{m_a} \sim 1 \frac{\text{cm}^2}{\text{g}} \; .
    \end{eqnarray}
    

\subsection{Asymmetric Dark Matter}
    
The idea of \textit{asymmetric} DM, formerly known as \textit{Technocosmology} \cite{Nussinov:1985xr}, is that there exists a new dark baryon quantum number $B_D$ in addition to the standard baryon number $B_V$, with DM being the lightest charged particle under $B_D$. We can define the following linear combination:  
    \begin{equation}
        B_{\rm conserved} = B_V - B_D \; ; \qquad B_{\rm broken} = B_V + B_D \; ,
    \end{equation}
which are conserved or broken at some energy scale. If the universe have a generalised global baryon symmetry the first combination is unbroken at all energy, while the second one is broken. At the breaking scale we have a net production of $\Delta B_{\rm broken}$ such that 
\begin{equation}
     \Delta B_V = \Delta B_D = \Delta B_{\rm broken}  / 2 \; .
\end{equation}
If the universe does not have a generalised global baryon symmetry, instead some asymmetry is generated either on $B_D$ or $B_V$. The mass of such asymmetric DM candidate is \cite{Harvey:1990qw}:
\begin{equation}
    m_{DM} = m_p \frac{\Omega_{DM}}{\Omega_b} \frac{\eta \left(B_V\right)}{\eta \left(B_D\right)/q_{DM}} \; ,
\end{equation}
where the ratios $\eta$ in the right hand side of the above equation are the charge-to-entropy ratios. For a baryon-symmetric universe we have: 
\begin{eqnarray}
    m_{DM} = q_{DM} \times\left(1.6 - 5 \right) \text{GeV} \; ,\\
    \eta\left(B_D\right)/\eta\left(B_V \right) \sim \exp(-m_{DM}/T_D)\; ,
\end{eqnarray}
if the chemical decoupling of the two sectors happens when DM is non-relativistic.


\subsection{Minimality}

With minimality we mean a theory which contains a minimal number of ingredients, e.g. heavy neutrinos. SU(2) neutrinos are excluded by direct detection, unless we introduce a suppressed coupling to the Z boson. Another possibility is to consider a minimal number of new fields, e.g. a real scalar singlet. In this case \cite{Feng:2014vea} the Lagrangian becomes:
\begin{equation}
    \mathcal{L} = \mathcal{L}_{SM} + \frac{\left(\partial_{\mu}S\right)^2}{2} - \frac{b_2}{2}S^2 - \frac{a_2}{2}H^{\dagger}HS^2 - \frac{b_4}{4}S^4\;, 
\end{equation}
and:
\begin{equation}
    \langle S\rangle = 0 \; , \qquad \sigma_{SI}= \frac{a_2^2M_N^4f^2}{\pi m_s^2m_h^4} \; .
\end{equation}
The ``original'' minimal DM model (MDM) is given by an $SU(2)_L$ multiplet with a given spin and mass, see Ref. \cite{Cirelli:2005uq}:
\begin{eqnarray}
     \mathcal{L} =   \bar{\chi }\left(i\slashed{D} + M\right)\chi \; , \qquad \text{fermion} \; ,\\
    \mathcal{L} = |D_{\mu} \chi|^2 - M^2 |\chi|^2 \; , \qquad \text{scalar} \; .
\end{eqnarray}
We have to require vanishing hypercharge and electric charge equal to:
\begin{equation}
    Q = T_3 + Y\;,
\end{equation}
This implies that no operators could mediate decay (i.e. terms of type $\chi e H$, $\rightarrow \, \chi = eH$) and we are left with a very constrained set of possibilities: $n = 5 \, M = 9.4 TeV$ spin $1/2$ fermion. Such heavy $SU(2)$ particles have important Sommerfeld enhancement effects, affecting indirect DM constraints.

Other minimal possibility is to consider an inert doublet model; e.g. a new SU(2) Higgs doublet with $Z_2$ symmetry \cite{Deshpande:1977rw}. It turns out that lightest $Z_2$-odd particle is stable and a good WIMP.


\subsection{Dark Photons}

Another interesting possibility is given by \textit{dark photons}, see for example Ref. \cite{Essig:2013lka}. Here we have gauge bosons associated with a new U(1)$^\prime$ symmetry, with a kinetic mixing term:
\begin{equation}
    -\frac{\epsilon}{2\cos{\theta_W}}B_{\mu\nu}F^{'\mu\nu} \; , \qquad \epsilon e A^{' \mu}J^{\mu}_{EM} \; ,
\end{equation}
\begin{equation}
m_{A'} \ll  2m_e \; , \qquad \text{can be long-lived enough}
\end{equation}
These are produced in oscillation or processes like:
\begin{equation}
    \gamma + e^{\pm} \rightarrow A' + e^{\pm} \; \quad e^+ + e^- \rightarrow A' + \gamma \; ,
\end{equation}
or from dark photon condensates, (similar to the axion misalignment mechanism): 
\begin{equation}
    \Omega_{A'} \sim 0.3 \sqrt{\frac{m_{A'}}{1 \text{ keV}}}\left(\frac{H_{inf}}{10^{12} \text{ GeV}}\right) \; .
\end{equation}
If new dark photons are charged under a non-abelian symmetry group, kinetic mixing term is automatically prohibited, and the particle is stable; the Lagrangian is of type: 
\begin{equation}
    \mathcal{L} = \mathcal{L}_{SM} -\frac{1}{4} F^{'\mu\nu}F^{'}_{\mu \nu} + \left(D_{\mu} \phi \right)^{\dagger}\left(D^{\mu} \phi \right) - \lambda|\phi|^2 |H|^2 - \mu_{\phi}^2 |\phi|^2 - \lambda_{\phi}|\phi|^4 \; ,
    \end{equation}
    and has interesting phenomenology: e.g. for $SU(3)$ we have semi-annihilation and bright gamma ray lines ($\sim \alpha$ instead of $ \sim \alpha^2$, $\psi_i \psi_j \rightarrow \psi_k\gamma $).


\section{Conclusions}

The book of physics contains several blank pages when it comes to the fundamental nature of DM as a particle. The discovery of the particle nature of DM will likely offer a portal on new physics beyond the Standard Model, and opportunities to understand how the universe on large scales is shaped by the microscopic properties of DM. It is important for the scholar who embarks in the study of particle DM to be keenly aware of the scope and nature of work that others have produced thus far; especially useful it is to be familiar with the ``bag of tricks'' that scholars have accumulated so far and to be able to master the key tools of this trade. These notes are meant to help with this critical and exciting task.

\bibliographystyle{apalike}
\bibliography{Profumo}

\end{document}